%% file: draft14.tex
\documentclass[prd,showpacs,superscriptaddress]{revtex4}

\usepackage{epsfig}
\usepackage{latexsym}
\usepackage{graphicx}
\usepackage{bm}
\usepackage{color}
\usepackage{amsmath}
\usepackage{amssymb}

\input{user_def.tex} 

\begin{document}

\preprint{Edinburgh 2008/19, RBRC-728, BNL-HET-08/13}

\title{\Large\bf Proton lifetime bounds from chirally symmetric lattice QCD}

\author{Y.~Aoki}
\affiliation{RIKEN-BNL Research Center, Brookhaven National Laboratory, Upton, NY 11973, USA.}
\author{P.~Boyle}
\affiliation{SUPA, School of Physics, The University of Edinburgh, Edinburgh EH9 3JZ, UK}
\author{P.~Cooney}
\affiliation{SUPA, School of Physics, The University of Edinburgh, Edinburgh EH9 3JZ, UK}
\author{L.~Del Debbio}
\affiliation{SUPA, School of Physics, The University of Edinburgh, Edinburgh EH9 3JZ, UK}
\author{R.~Kenway}
\affiliation{SUPA, School of Physics, The University of Edinburgh, Edinburgh EH9 3JZ, UK}
\author{C.M.~Maynard}
\affiliation{EPCC, School of Physics, The University of Edinburgh, Edinburgh EH9 3JZ, UK}
\author{A.~Soni}
\affiliation{High Energy Theory Group, Brookhaven National Laboratory, Upton, NY 11973, USA}
\author{R.~Tweedie}
\affiliation{SUPA, School of Physics, The University of Edinburgh, Edinburgh EH9 3JZ, UK}

\collaboration{RBC--UKQCD Collaboration}
\noaffiliation{RBC--UKQCD Collaboration}


\date{\today}

\begin{abstract}
We present results for the matrix elements relevant for proton decay
in Grand Unified Theories (GUTs). The calculation is performed at a
fixed lattice spacing $a^{-1}=1.73(3)$~GeV using 2+1 flavors of domain
wall fermions on lattices of size $16^3\times32$ and $24^3\times64$
with a fifth dimension of length 16.  We use the indirect method which
relies on an effective field theory description of proton decay, where
we need to estimate the low energy constants, $\alpha =
-0.0112(25)$~${\rm{GeV^3}}$ and $\beta = 0.0120(26)$~${\rm{GeV^3}}$.
We relate these low energy constants to the proton decay matrix
elements using leading order chiral perturbation theory.  These can
then be combined with experimental bounds on the proton lifetime to
bound parameters of individual GUTs.

\end{abstract}

\maketitle

\section{Introduction}
\label{sec:intro}

Proton decay is a distinctive experimental signature of Grand Unified
Theories (GUTs). Decay experiments can test the predictions of these
theories, and, even though direct nucleon decays have not been
observed, the experimental lower bound on the decay rate has already
ruled out the simplest minimal supersymmetric
models~\cite{Murayama:2001ur}. One of the expected decay channels is
$N \rightarrow \mathrm{M} + l$, where $N$ and $\mathrm{M}$ indicate
respectively the nucleon and a pseudoscalar meson ($K, \pi$), while
$l$ is a lepton ($e$, $\mu$, $\nu_e$, $\nu_\mu$). This decay is
induced by supersymmetric particles or heavy gauge boson exchange,
which can be integrated out to obtain an effective Lagrangian
describing the low--energy dynamics in terms of the usual Standard
Model fields.  The lowest--dimensional operators that appear in this
approach are $(\bar q^c q)(\bar l^c q)$ operators of dimension six;
the transition amplitude for the decay is proportional to their
hadronic matrix elements $\langle M | (\bar q^c q)(\bar l^c q) |
N\rangle$. A quantitative estimate of such hadronic matrix elements,
which requires taming non--perturbative QCD effects, is a key
ingredient in probing the effects of higher--dimensional operators in
GUTs models at current experiments.

Several determinations of the hadronic matrix elements have been
performed in the
past~\cite{Ioffe:1981kw,Tomozawa:1980rc,Donoghue:1982jm,
Meljanac:1981xd,Krasnikov:1982gf,Thomas:1983ch,Ioffe:1983ju,Brodsky:1983st,
Hara:1986hk,Bowler:1987us,Gavela:1988cp,Aoki:1999tw,Tsutsui:2004qc,
Aoki:2006ib}, using either QCD bound state phenomenological models, or
lattice QCD. Due to recent progress in simulating dynamical fermions,
lattice QCD has become a quantitative method to compute hadronic
matrix elements from first principles with controlled systematic
uncertainties.  The matrix elements relevant for nucleon decay can be
extracted from three-point correlators computed on the lattice. This
is the so--called {\it direct}\ method, which requires an expensive
computer simulation. However, the same matrix elements can also be
computed, at the cost of introducing difficult to estimate
systematics, using the chiral lagrangian describing proton
decay~\cite{Claudson:1981gh}: in this
case they are expressed as functions of the low--energy constants
(LECs) that appear in the chiral lagrangian. These LECs can be
computed from two--point lattice correlators at a lesser computational
cost. However, such an {\it indirect}\ determination of the matrix
elements depends on the accuracy of chiral perturbation theory, and
therefore is affected by an additional source of systematic error.

In this work, we present a new determination of the matrix elements
that are relevant for nucleon decay based on the {\it indirect}\
method, using dynamical Domain Wall Fermion (DWF) configurations with
$2+1$ flavours. Our results extend the ones obtained for $2$ flavours
of dynamical DWF in Ref.~\cite{Aoki:2006ib}. Systematic errors are
greatly reduced by simulating at light quark masses, and using
non--perturbative renormalization. Note that the exponentially
suppressed chiral symmetry breaking of DWF greatly simplifies the
mixing of operators under renormalization, which improves the
precision of the final result.  The correct number of flavours gives
confidence in setting the scale, a large source of uncertainty in some
early lattice determinations.

The chiral perturbation theory results used for this work are
summarized in Section~\ref{sec:chpt}, which also sets the notation
used throughout the paper. The chiral Lagrangian for nucleon decay
involves two LECs, which are obtained by extrapolating to the chiral
limit the outcome of numerical simulations performed at light quark
masses. A {\it direct}\ measurement of the hadronic matrix element
using lattice three--point functions, which relies much less on the
validity of chiral perturbation theory, is in progress, and is
deferred to a subsequent publication.

Details of our lattice simulations are reported in
Section~\ref{sec:latt}. Our results are obtained from gauge
configurations with volumes of $16^3\times32$ and $24^3\times64$.
Both have a fifth dimension of size $L_s = 16$ and use the Iwasaki
gauge action. The lattice spacing is $a\approx 0.114~\mathrm{fm}$
corresponding to physical volumes of $(1.8~\mathrm{fm})^3$, and
$(2.7~\mathrm{fm})^3$ respectively. Being able to compare two
different physical volumes enables us to estimate the finite volume
effects. Results for meson spectroscopy and topology have already been
presented in Ref.~\cite{Antonio:2007mh} for the smaller lattice and in
Ref. ~\cite{Allton:2008pn} for the larger lattice. We refer to these
publication for details of calculations involving the lattices which
are used in this paper.  The range of fermion masses simulated for
this work yields a ratio of the pseudoscalar to vector meson masses in
the range $0.378 \leq m_\mathrm{PS}/m_\mathrm{V} \leq 0.615$. Working
at fixed lattice spacing, we are not able to present a continuum
extrapolation for our final result. Nonetheless, it should be noted
that DWFs are automatically O(a) improved \footnote{Due
to finite $L_s$ there are small $\mathcal{O}(a)$ errors, which, however are
negligible compared to the other errors in this study},
and is therefore expected to have a good scaling behavior.

Section~\ref{sec:NPR} presents our results for the non--perturbative
renormalization (NPR) of the three--quark operators, using the RI--MOM
scheme. We compute the renormalization mixing matrix and perform the
matching required to obtain the renormalized operators in the
$\overline{\mathrm{MS}}$-scheme.

In the last Section of the paper, we combine the lattice amplitudes
with the renormalization factors to compute the phenomenologically
relevant matrix elements in the $\overline{\mathrm{MS}}$ scheme. We
discuss the error budget in detail including estimates of the
systematic error due to the chiral extrapolation, the renormalization,
the finite volume and the choice of method to set the lattice spacing,
and the foreseeable improvements on the current estimate.

\section{Chiral lagrangian for proton decay}
\label{sec:chpt}

Integrating out the heavy GUT particles yields the low--energy
effective Lagrangian describing nucleon decay written in terms of the
QCD fundamental fields:
\begin{equation}
\label{eq:deltaBlag}
{\mathcal L}^{\Delta B}=\sum_{d=1,2}\sum_{i=1}^4 C^{(i)}_d
Q^{(i)}_d + \sum_{d=1,2} \sum_{i=1}^6\tilde C^{(i)}_d
\tilde Q^{(i)}_d,
\end{equation}
where $d$ denotes the generation of the lepton produced in the decay,
and $i$ is a label for the dimension--six operators containing three
quark and one lepton field that describe nucleon decay. $C^{(i)}_d$
and $\tilde C^{(i)}_d$ are Wilson coefficients. The full list of
operators $Q^{(i)}$, $\tilde Q^{(i)}$ was identified on symmetry
grounds in Refs.~\cite{Weinberg:1979sa,Wilczek:1979hc,Abbott:1980zj};
their matrix elements between hadronic states determine the decay
amplitude. For instance, the matrix elements that are relevant for the
process where a proton decays into a pion are:
\begin{equation}
\label{eq:matel}
\langle \pi(\vec{p}) | \epsilon_{abc} (u^{aT} CP_{R,L} d^{b}) P_L u^{c}|p(\vec{k},s)\rangle 
= P_L\left[W_0^{R/L L}(q^2) - i\slsh{q}W_q^{R/L L}(q^2) \right]u(k,s),
\end{equation}
where $a,b,c$ are colour indices, $C$ is the charge-conjugation
operator, and $P_{R,L}=\frac{1\pm\gamma_5}{2}$ are the right-- and
left--handed projectors. The non--perturbative dynamical effects are
captured by the two form factors that appear on the RHS of
Eq.~(\ref{eq:matel}), while $q$ (the momentum carried by the electron)
is the momentum transfer.  It is convenient to introduce here a
generic notation for three--quark operators with an arbitrary spin
structure:
\begin{equation}
\mathcal{O}^{\Gamma\Gamma^{\prime}}(\vec{x},t)
        = \epsilon_{abc} \left[ u^{a}(\vec{x},t) (C\Gamma)
        d^{b}(\vec{x},t) \right] \Gamma^{\prime} u^{c}(\vec{x},t).
\end{equation}
where $\Gamma$ and $\Gamma^\prime$ are elements of the Clifford
algebra in four--dimensional Euclidean spacetime, and we have omitted
spinor indices. We use the notation $S=1$, $P=\gamma_5$,
$V=\gamma_\mu$, $A=\gamma_{\mu}\gamma_5$,
$T=\sigma_{\mu\nu}=\frac{1}{2}\{\gamma_\mu,\gamma_\nu\}$, $R=P_R$, and
$L=P_L$. Further operators with this structure appear when computing
the nucleon mass, and upon renormalization, as discussed in
Sects.~\ref{sec:latt}, and~\ref{sec:NPR}.

Following the notation in Refs.~\cite{Claudson:1981gh,Aoki:1999tw},
the chiral Lagrangian describing baryon--meson dynamics is written in
terms of a pseudoscalar meson (octet) field:
\begin{equation}
\phi=\left( 
\begin{array}{ccc}
\sqrt{\frac12}\pi^0+\sqrt{\frac16}\eta & \pi^+ & K^+ \\
\pi^- & -\sqrt{\frac12}\pi^0+\sqrt{\frac16}\eta & K^0 \\
K^- & \bar K^0 & -\sqrt{\frac23}\eta \\
\end{array}
\right),
\end{equation}
and a spinor baryon (octet) field:
\begin{equation}
B=\left( 
\begin{array}{ccc}
\sqrt{\frac12}\Sigma^0+\sqrt{\frac16}\Lambda^0 & \Sigma^+ & p \\
\Sigma^- & -\sqrt{\frac12}\Sigma^0+\sqrt{\frac16}\Lambda^0 & n \\
\Xi^- & \Xi^0 & -\sqrt{\frac23}\Lambda^0 \\
\end{array}
\right).
\end{equation}
At lowest order in powers of momentum, and in Euclidean space-time,
the chirally symmetric Lagrangian is written as:
\begin{align}
{\mathcal L}_0 = & \frac{f^2}{8} \Tr (\partial_\mu
\Sigma)(\partial_\mu \Sigma^\dagger) + \Tr \bar B (\gamma_\mu
\partial_\mu + M_B) B \nonumber \\
&+ \frac12 \Tr \bar B \gamma_\mu \left[\xi\partial_\mu \xi^\dagger +
\xi^\dagger \partial_\mu \xi \right] B + \frac12 \Tr \bar B \gamma_\mu
B
\left[\left(\partial_\mu\xi\right)\xi^\dagger +
  \left(\partial_\mu\xi^\dagger\right)\xi \right] \nonumber \\
& - \frac12 (D-F) \Tr \bar B \gamma_\mu \gamma_5 B 
\left[\left(\partial_\mu\xi\right)\xi^\dagger -
  \left(\partial_\mu\xi^\dagger\right)\xi \right] \nonumber \\
& + \frac12 (D+F) \Tr \bar B \gamma_\mu \gamma_5 
\left[\xi \partial_\mu\xi^\dagger -
  \xi^\dagger \partial_\mu\xi \right] B,
\end{align}
where the unitary matrices $\Sigma$ and $\xi$ are defined as:
\begin{equation}
\Sigma = \exp\left(\frac{2 i \phi}{f}\right),
~~~~~\xi = \exp\left(\frac{i \phi}{f}\right) .
\end{equation}
Introducing a diagonal quark mass matrix:
\begin{equation}
M=\left(
\begin{array}{ccc}
m_u & & \\
& m_d & \\
& & m_s
\end{array}
\right),
\end{equation}
the symmetry--breaking part of the chiral Lagrangian becomes:
\begin{align}
{\mathcal L}_1 =& -v^3 \Tr \left(\Sigma^\dagger M + M \Sigma\right)
-a_1 \Tr \bar B \left(\xi^\dagger M \xi^\dagger +\xi M \xi\right) B
-a_2 \Tr \bar B B \left(\xi^\dagger M \xi^\dagger +\xi M \xi\right) \nonumber \\
& -b_1 \Tr \bar B \gamma_5 \left(\xi^\dagger M \xi^\dagger -\xi M \xi\right)
B 
-b_2 \Tr \bar B \gamma_5 B \left(\xi^\dagger M \xi^\dagger -\xi M \xi\right).
\end{align}
The low--energy constants that appear in the chiral Lagrangian are
extracted from phenomenological analyses. In particular, following the
notation in Ref.~\cite{Claudson:1981gh}, $f$ is the pion decay
constant in the chiral limit, 130(5) MeV \cite{PDBook}. The
combination $F+D$ yields the nucleon axial charge,
$g_A=1.2695(29)$~\cite{PDBook}, while the combination $F-D$ is
related to the ratio of the zero--momentum form factors for
semileptonic hyperon decay, $g_1/f_1$~~\cite{Hsueh:1988ar}. Together
these give, $F=0.47(1)$ and $D=0.80(1)$. $a_1$ and $a_2$ are
symmetry--breaking parameters, but their values are not required in
this work. The parameters $b_1$,$b_2$ are not precisely determined and
are an extra source of systematic error.

The transformation properties under SU(3)$_L \times$ SU(3)$_R$ of the
three-quark operators in Eq.~\ref{eq:deltaBlag} determine the
expression of the baryon--number violating operators in the chiral
Lagrangian. The latter appear in the Lagrangian with two new
low--energy constants $\alpha$ and $\beta$~\cite{Claudson:1981gh}:
\begin{align}
\label{eq:deltaBchlag}
{\mathcal L}^{\Delta B}=
& \alpha \sum_{d=1}^{2} \left\{ C_{d}^{(1)}
\left[ e_{dL}\Tr {\mathcal F} \xi B_L\xi - \nu_{dL} \Tr {\mathcal
F}^{\prime} \xi B_L\xi \right]\right.\nn\\ 
& + C_{d}^{(2)}
e_{dR}\Tr {\mathcal F} \xi^{\dagger} B_R \xi^{\dagger} +
\tilde{C}_{d}^{(1)} \left[ e_{dL}\Tr \tilde{{\mathcal F}} \xi B_L\xi -
\nu_{dL} \Tr \tilde{{\mathcal F}}^{\prime} \xi B_L\xi \right]\nn\\ 
& \left. + \tilde{C}_{d}^{(2)} e_{dR}\Tr \tilde{{\mathcal F}}
\xi^{\dagger} B_R \xi^{\dagger} + \tilde{C}_{d}^{(5)} \nu_{dL} \Tr
\tilde{{\mathcal F}}^{\prime\prime} \xi B_L\xi \right\} + \nn\\ &  \beta
\sum_{d=1}^{2} \left\{ C_{d}^{(3)} \left[ e_{dL}\Tr {\mathcal F} \xi
B_L\xi^{\dagger} - \nu_{dL} \Tr {\mathcal F}^{\prime} \xi
B_L\xi^{\dagger} \right]\right. \nn\\ 
& + C_{d}^{(4)} e_{dR}\Tr
{\mathcal F} \xi^{\dagger} B_R \xi + \tilde{C}_{d}^{(3)} \left[
e_{dL}\Tr \tilde{{\mathcal F}} \xi B_L\xi^{\dagger} - \nu_{dL} \Tr
\tilde{{\mathcal F}}^{\prime} \xi B_L\xi^{\dagger} \right] \nn\\ 
& \left .+ \tilde{C}_{d}^{(4)} e_{dR}\Tr \tilde{{\mathcal F}}
\xi^{\dagger} B_R \xi + \tilde{C}_{d}^{(6)} \nu_{dL} \Tr
\tilde{{\mathcal F}}^{\prime\prime} \xi B_L\xi^{\dagger} \right\} +
{\rm{h.c.}}
\end{align}
The matrices ${\mathcal F},{\mathcal F}^{\prime},\tilde{{\mathcal
F}},\tilde{{\mathcal F}}^{\prime}$, and $\tilde{{\mathcal
F}}^{\prime\prime}$ are projectors in flavour space; their explicit
expressions are:
\be
\label{eq:proj}
{\mathcal F} = \lf(\begin{array}{ccc}
  0 & 0 & 0\\
  0 & 0 & 0\\
  1 & 0 & 0\end{array}\rr),
{\mathcal F}^{\prime} = \lf(\begin{array}{ccc}
  0 & 0 & 0\\
  0 & 0 & 0\\
  0 & 1 & 0\end{array}\rr),
\tilde{{\mathcal F}} = -\lf(\begin{array}{ccc}
  0 & 0 & 0\\
  1 & 0 & 0\\
  0 & 0 & 0\end{array}\rr),
\tilde{{\mathcal F}}^{\prime} = -\lf(\begin{array}{ccc}
  0 & 0 & 0\\
  0 & 1 & 0\\
  0 & 0 & 0\end{array}\rr),
\tilde{{\mathcal F}}^{\prime\prime} = \lf(\begin{array}{ccc}
  0 & 0 & 0\\
  0 & 0 & 0\\
  0 & 0 & 1\end{array}\rr).
\ee

Eqs.~\ref{eq:deltaBchlag} and~\ref{eq:proj} show that the
low--energy constants $\alpha$ and $\beta$ determine the matrix
elements:
\begin{align}
\label{eq:alpha_def}
\langle 0 | \mathcal{O}^{RL} | p({\mathbf k},s) \rangle=& \alpha\ 
P_L u({\mathbf k},s) \\
\label{eq:beta_def}
\langle 0 | \mathcal{O}^{LL} | p({\mathbf k},s) \rangle=& \beta\ 
P_L u({\mathbf k},s)
\end{align}
where $u({\mathbf k},s)$ is the spinor associated with a proton of
momentum ${\mathbf k}$ and spin projection $s$. The phase definition
is fixed such that $\alpha$ and $\beta$ are real and $\alpha<0$. As we
will later describe, we observe $\alpha+\beta\simeq 0$, which is
expected because of the relation,
\begin{equation}
 (\alpha+\beta)\ u({\mathbf k},s) = -\langle 0|
  \epsilon^{abc} (u^{T a} C d^b) \gamma_5u^c  | 
  p({\mathbf k},s)\rangle,
\end{equation}
which vanishes in the non-relativistic limit and is known to be quite
small even at small quark masses~\cite{Sasaki:2001nf}. 

Using chiral perturbation theory to compute the matrix element in
Eq.~(\ref{eq:matel}) yields for the $N \rightarrow \pi$
transition~\cite{Claudson:1981gh,Aoki:1999tw}:
\ba
\bra{\pi^0}\mathcal{O}^{RL}\ket{p({\mathbf k},s)} 
&=& \alpha P_L u({\mathbf k},s) 
\lf[ \frac{1}{\sqrt{2}f} - \frac{D+F}{\sqrt{2}f} 
\frac{-q^2+m_N^2}{-q^2-m_N^2} - \frac{4b_1}{\sqrt{2}f}
\frac{m_u m_N}{-q^2-m_N^2} \rr]\nn\\
&-& \alpha P_L i \lslash{q} u({\mathbf k},s) \lf[
  \frac{D+F}{\sqrt{2}f} \frac{2m_N}{-q^2-m_N^2} +
  \frac{4b_1}{\sqrt{2}f}
\frac{m_u}{-q^2-m_N^2} \rr],\\
\bra{\pi^0}\mathcal{O}^{LL}\ket{p({\mathbf k},s)} 
&=& \beta P_L u({\mathbf k},s) \lf[ \frac{1}{\sqrt{2}f} - 
\frac{D+F}{\sqrt{2}f} \frac{-q^2+m_N^2}{-q^2-m_N^2} - 
\frac{4b_1}{\sqrt{2}f}\frac{m_u m_N}{-q^2-m_N^2} \rr]\nn\\
&-& \beta P_L i \lslash{q} u({\mathbf k},s) \lf[ \frac{D+F}{\sqrt{2}f} 
\frac{2m_N}{-q^2-m_N^2} + \frac{4b_1}{\sqrt{2}f}\frac{m_u}{-q^2-m_N^2} \rr],
\ea
where $q$ is the four-momentum of the outgoing lepton. 
In the limit where $q^2\ll m_N^2$ and $b_1 m_u \ll m_N$, these expressions simplify to:
\ba
\bra{\pi^0}\mathcal{O}^{RL}\ket{p({\mathbf k},s)} 
&\simeq& \alpha P_L u({\mathbf k},s) \lf[\frac{1}{\sqrt{2}f} + 
\frac{D+F}{\sqrt{2}f} \rr] + O(m_l^2/m_N^2), \label{eq:me1}\\
\bra{\pi^0}\mathcal{O}^{LL}\ket{p({\mathbf k},s)} 
&\simeq& \beta P_L u({\mathbf k},s) \lf[\frac{1}{\sqrt{2}f} + 
\frac{D+F}{\sqrt{2}f}  \rr] + O(m_l^2/m_N^2),\label{eq:me2}
\ea
where $-q^2=m_l^2$ is the on--shell condition for the outgoing
lepton. The equations above relate the proton decay matrix elements to
the low--energy constants $\alpha$ and $\beta$; note that, in order to
reconstruct the matrix elements on the {\it lhs}\ of Eqs.~\ref{eq:me1}
and~\ref{eq:me2} using the indirect method, the combination
$F+D$ and the pion decay constant, $f$, are also required.

\section{Lattice simulations}
\label{sec:latt}

\subsection{Dataset description}

The analysis was performed on 2+1 flavor DWF ensembles with volumes of
$16^3 \times 32$ and $24^3 \times 64$ generated using the Iwasaki
gauge action with $\beta=2.13$ and the domain wall fermion quark
action with $L_s = 16$. At each volume we generated sets of
configurations with a light isodoublet with masses $am_{ud}$ = 0.005
($24^3 \times 64$ only), 0.01, 0.02 or 0.03 and a fixed approximate
strange quark mass, $am_s = 0.04$. The ensembles, described
in~\cite{Allton:2007hx} and \cite{Allton:2008pn}, have a fixed inverse
lattice spacing of $a^{-1} = 1.73(3)$~GeV and were generated with the
RHMC algorithm with a trajectory length of $\tau=1$. These same
datasets were used to calculate $g_A$, \cite{Yamazaki:2008py}. The
configurations used for both the non--perturbative renormalisation and
the matrix element calculation are shown in Table~\ref{tab:datasets}.

\input{datasets.tex}

For each of the seven ensembles matrix elements were calculated using
correlation functions composed of valence quarks with masses equal to
the light quark mass in the sea. To improve statistics, correlators
were oversampled and averaged into bins whose size depended on the
Monte Carlo time separation between measurements. The binning was
consistent with the integrated auto-correlation length for the
pseudoscalar meson correlators at the time separation typically used,
which was calculated to be of order 50 trajectories. Multiple sources
per configuration and several different types of smearing have also
been used to improve the signal. As well as local sources ($L$), we
employ gauge--invariant Gaussian smearing with two different smearing
radii ($G$ and $G*$) and gauge fixed hydrogen--like wavefunction
smearing ($H$). One or both of the propagators used to construct the
two-point correlators for mesons may be smeared while for baryons,
one, two or all three propagators may be smeared. We adopt the same
convention used in Ref.~\cite{Antonio:2006px} for naming the smeared
two--point functions.

The chiral limit is defined as the value of $am_f$ such that $am_f +
am_{\mathrm{res}} = 0$, where $am_{\mathrm{res}} = 0.00315(2)$ is the
residual quark mass, estimated in
Refs.~\cite{Antonio:2006px,Allton:2006ax,Allton:2007hx}. The lattice
scale is determined from a combination of the $\Omega^-$ baryon mass
and the pseudoscalar kaon and pion masses, yielding a value $a^{-1} =
1.729(28)$ GeV (see Ref. ~\cite{Allton:2008pn} for details).

The parameters used in the simulations correspond to a pseudoscalar
meson mass ranging from $331$ MeV to $671$ MeV. The renormalization
constant for the axial current, $Z_A=0.7162(2)$, which we will use in
the non-perturbative renormalization of the nucleon decay operators,
was obtained from a hadronic matrix element of the conserved DWF axial
current in Ref.~\cite{Allton:2006ax}.

\subsection{Nucleon mass and amplitude}
\label{subsec:nucmassamp}
\begin{figure}
{\includegraphics[width=.32\textwidth, bb = 0 0 520 490]{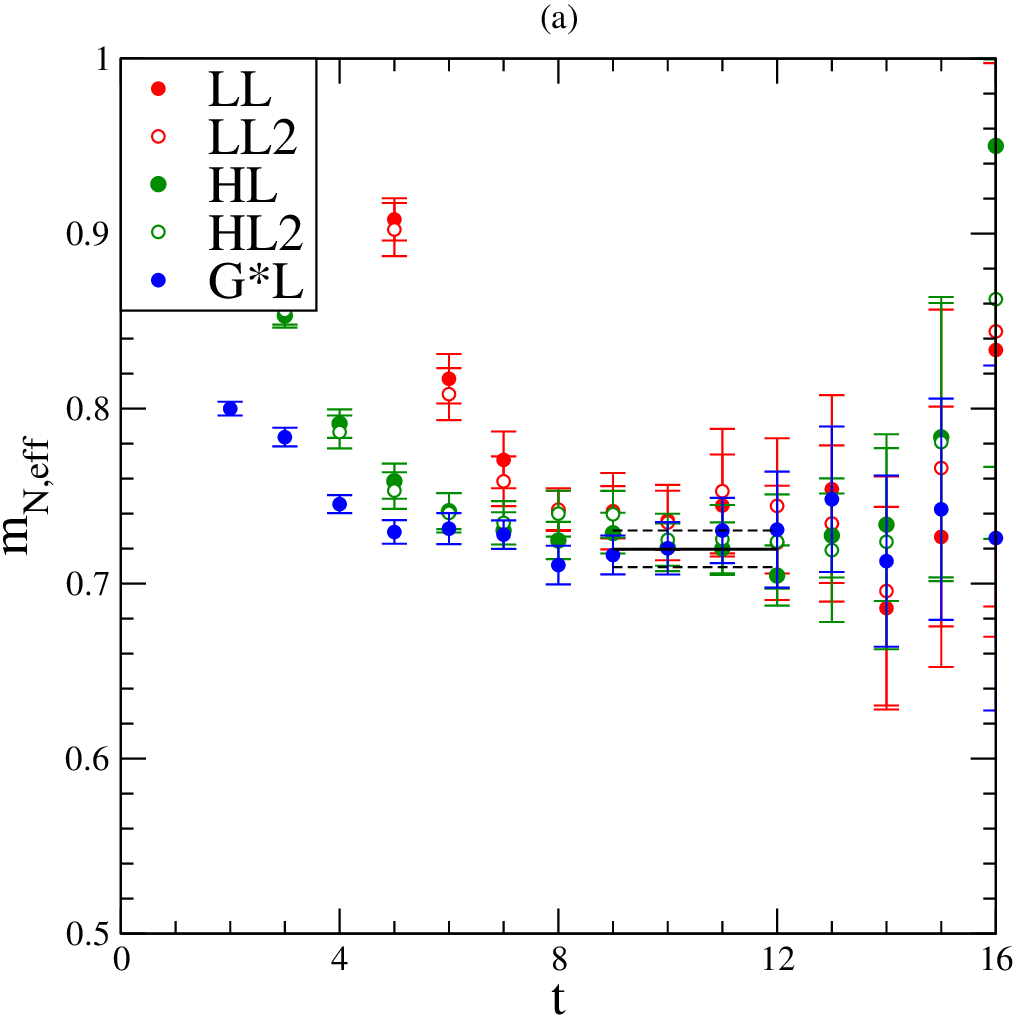}}
{\includegraphics[width=.32\textwidth, bb = 0 0 520 490]{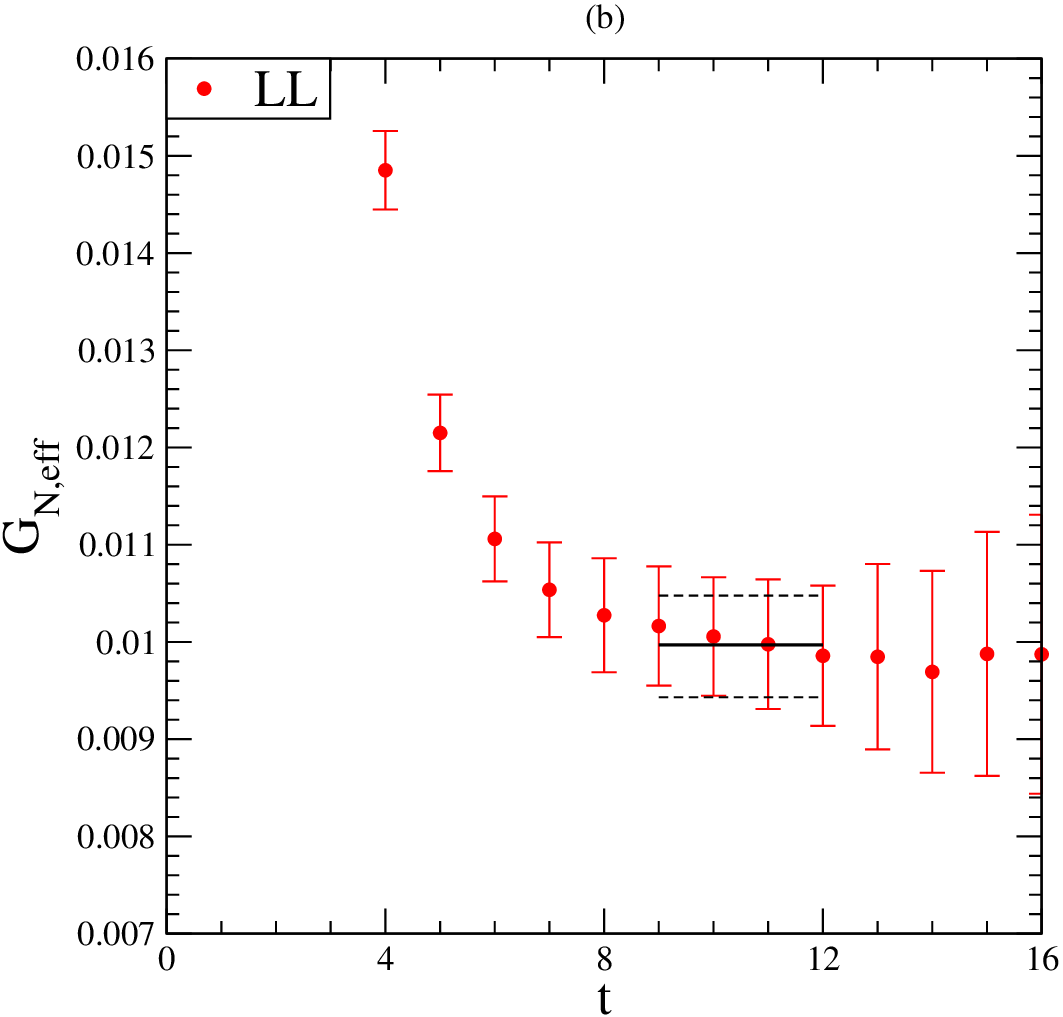}}
{\includegraphics[width=.32\textwidth, bb = 0 0 520 490]{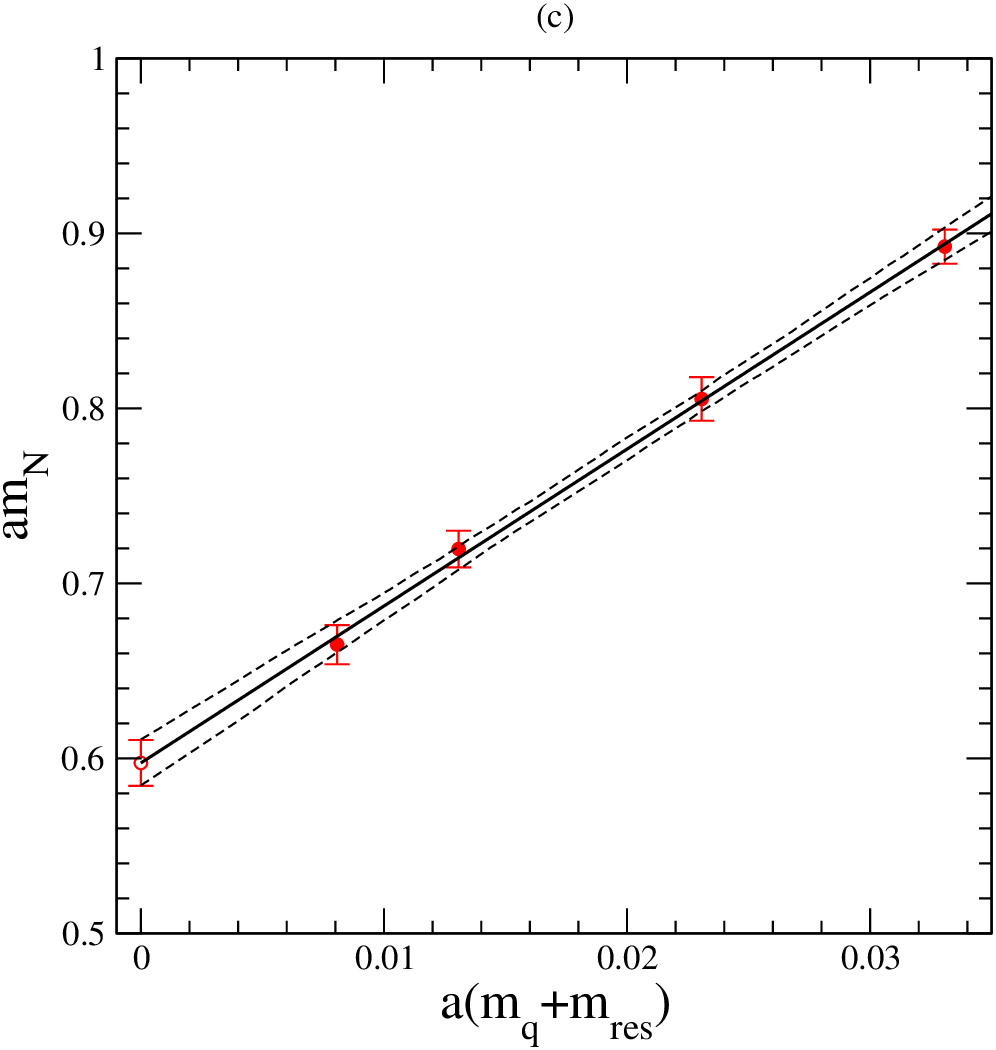}}
\caption{\label{fig:n-mass} (a) is an effective mass plot and (b) is
an effective amplitude (Eq.\ref{eq:geff}) plot for the nucleon. Both
are calculated on the $24^3 \times 64$ dataset with $am_u = 0.01$. The
different colours in the effective mass plot correspond to different
smearings. Datasets are labelled with the smearing (i.e. LL). Those
datasets labelled with a 2 use the operator $f_{A_4S,A_4S}(t)$, the
rest use $f_{PS,PS}(t)$. (c) is a linear extrapolation of the ground
state mass to the chiral limit.}
\end{figure}

Starting from the correlator of two operators,
$\mathcal{O}^{\Gamma_1\Gamma_2}$ and $\mathcal{O}^{\Gamma_3\Gamma_4}$,
we can define the scalar two--point function:
\begin{equation}
f_{\Gamma_1\Gamma_2,\Gamma_3\Gamma_4}(t)=\sum_{\vec{x}} \Tr\ 
\left[ \langle \mathcal{O}^{\Gamma_1,\Gamma_2}(\vec{x},t)
 \overline{\mathcal{O}}^{\Gamma_3,\Gamma_4}(0)\rangle 
\left(\frac{1+\gamma_4}{2}\right) \right].
\end{equation}
Using the notation introduced so far, $\mathcal{O}^{PS}(\vec{x},t)$ is
the usual local proton interpolating operator:
\begin{equation}
\mathcal{O}^{PS}(\vec{x},t) = \epsilon_{abc}
\left[u^{a T}(\vec{x},t) C\gamma_{5}d^{b}(\vec{x},t) \right] u^{c}(\vec{x},t),
\end{equation}
and the large--time exponential fall--off of the correlator $f_{PS,PS}$ is
dictated by the nucleon mass:
\begin{equation}
f_{PS,PS}(t) = 2 e^{-am_N t} G_N^2 + \ldots,
\end{equation}
where $G_N$ is the overlap of the proton interpolating field to
the normalized proton state:
\begin{equation}
 \langle 0 | \mathcal{O}^{PS}(\vec{0},0) | p({\mathbf k},s)\rangle = 
G_N u({\mathbf k},s).
\end{equation}
The nucleon mass is obtained from the two--point functions
$f_{PS,PS}(t)$ and $f_{A_4S,A_4S}(t)$,
each of them being computed for several smearing combinations.
Firstly, for each two--point function and for each smearing
combination, we calculate the effective mass:
\begin{equation}
\label{eq:meff}
m_{N,\mathrm{eff}}(t) = \log \left[ \frac{f(t)}{f(t+1)} \right]
\end{equation}
where $f(t)$ indicates the two--point function in any one of the
channels used for the analysis. Results for the effective mass
computed from both two--point functions, and for two different
smearing combinations, are reported in Fig.~\ref{fig:n-mass}. The
agreement between the different channels within the error bars is
clearly seen in the first plot on the left for the $24^3 \times 64$
data with $am_u=0.01$. The effective mass can be fitted to the same
constant $m$ for each channel; correlations between different
time--slices are taken into account by minimizing a correlated
$\chi^2$:
\begin{equation}
\chi^{(n)2}(m^{(n)}) = \sum_{t,t^\prime} \left[ m^{(n)}_{N,\mathrm{eff}}(t) - m^{(n)} \right]
C^{(n)-1}_{tt^\prime} \left[ m^{(n)}_{N,\mathrm{eff}}(t^\prime) - m^{(n)} \right]
\end{equation}
where $C^{(n)}_{tt^\prime}$ is the covariance matrix:
\begin{equation}
C^{(n)}_{tt^\prime} = \frac{1}{N_{\mathrm{boot}}}\sum_{m=1}^{N_\mathrm{boot}}
\left[\bar{m}_{N,\mathrm{eff}}^{(n,m)}(t) - \langle\bar{m}_{N,\mathrm{eff}}^{(n)}(t)\rangle \right]
\left[\bar{m}_{N,\mathrm{eff}}^{(n,m)}(t') - \langle\bar{m}_{N,\mathrm{eff}}^{(n)}(t')\rangle \right]
\end{equation}
the index $n$ represents a bootstrap resampling of the original
data, and the index $m$ represents a bootstrap resampling of the
$n^\mathrm{th}$ bootsample.  ${N_\mathrm{boot}}$ is the number of
bootstrap samples, $\bar{m}_{N,\mathrm{eff}}^{(n,m)}(t)$ is the
effective mass determined from the $m^\mathrm{th}$ resampling of the
$n^\mathrm{th}$ bootsample and
$\langle\bar{m}_{N,\mathrm{eff}}^{(n)}(t)\rangle$ is the average of
the effective mass over the $m^\mathrm{th}$ resampling of the
$n^\mathrm{th}$ bootsample.

All channels display a plateau for the effective mass, and the
limiting values are compatible within the statistical errors. The
smeared propagators reach the limiting value earlier, as expected,
thus yielding a longer plateau for the fit to be performed.  In order
to increase the precision of the fit, all channels were fitted
simultaneously to a single constant $m$; correlations between
different channels are also taken into account in the construction of
the covariance matrix. Note that with this fitting procedure several
channels are fitted simultaneously without adding extra parameters,
and the minimization in the one--dimensional parameter space can be
performed analytically.

The fit to the amplitude, $G_N$, is subsequently performed by defining
an effective amplitude:
\begin{equation}
\label{eq:geff}
G^2_{N,\mathrm{eff}}(t) = \frac12 f_{PS,PS}(t) \exp(m t),
\end{equation}
where $m$ is the nucleon mass obtained in the fit described above, and
we used the same notation as in Eq.~\ref{eq:meff} to denote the
two-point function. $G_N$ is then obtained from a fit to a constant by
minimizing a fully correlated $\chi^2$, as discussed above for the
nucleon mass. Results for $G_{N,\mathrm{eff}}(t)$ are displayed in
Fig.~\ref{fig:n-mass}(b), where a long plateau is clearly visible.

For all the fits presented here, the results of the minimization
procedure are stable with respect to sensible variations of the fit
range. All the correlators, smearings, and fit ranges are summarized
in Table~\ref{tab:fitranges}. Variations of the fitted parameters
remain within their statistical error as the bounds of the fitting
range are shifted by $\pm 1$ timeslice.

\begin{figure}
\includegraphics[width=.4\textwidth, bb = 0 0 520 490]{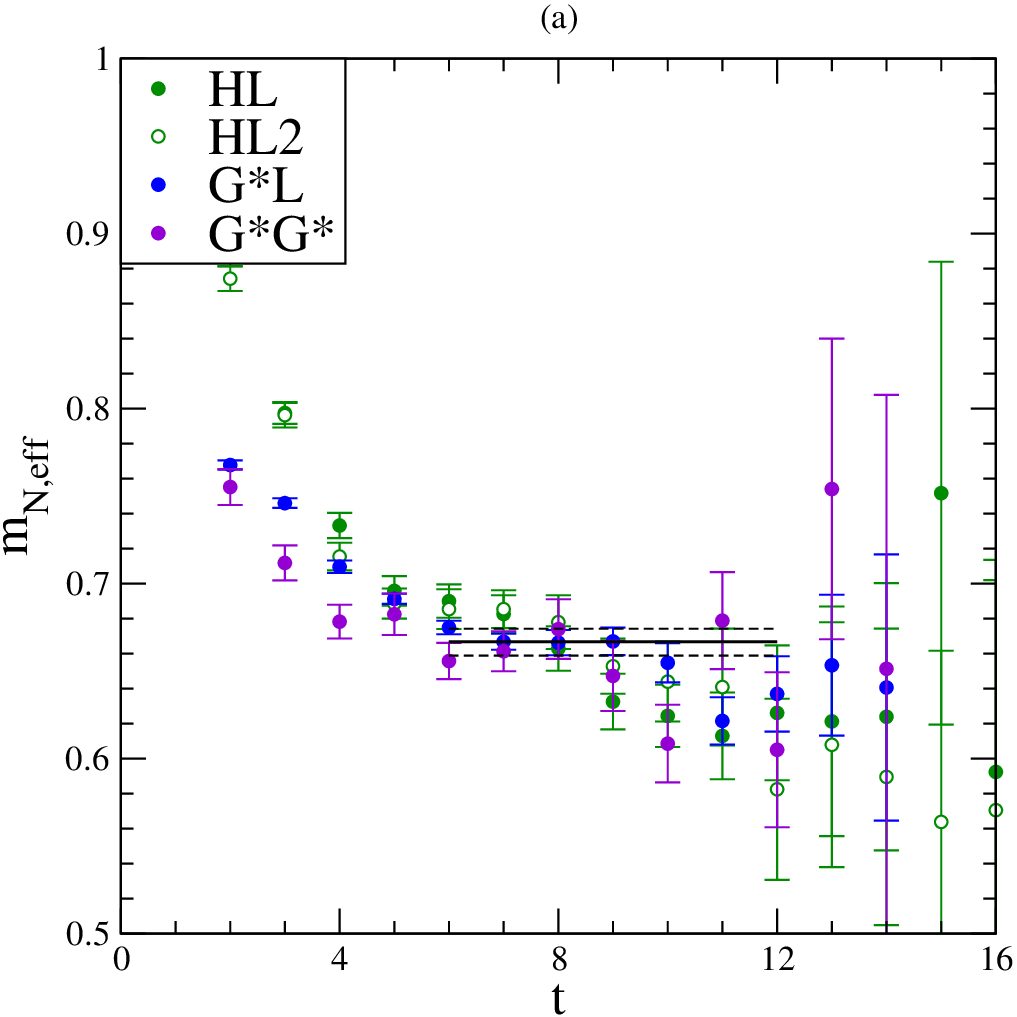}
\includegraphics[width=.4\textwidth, bb = 0 0 520 490]{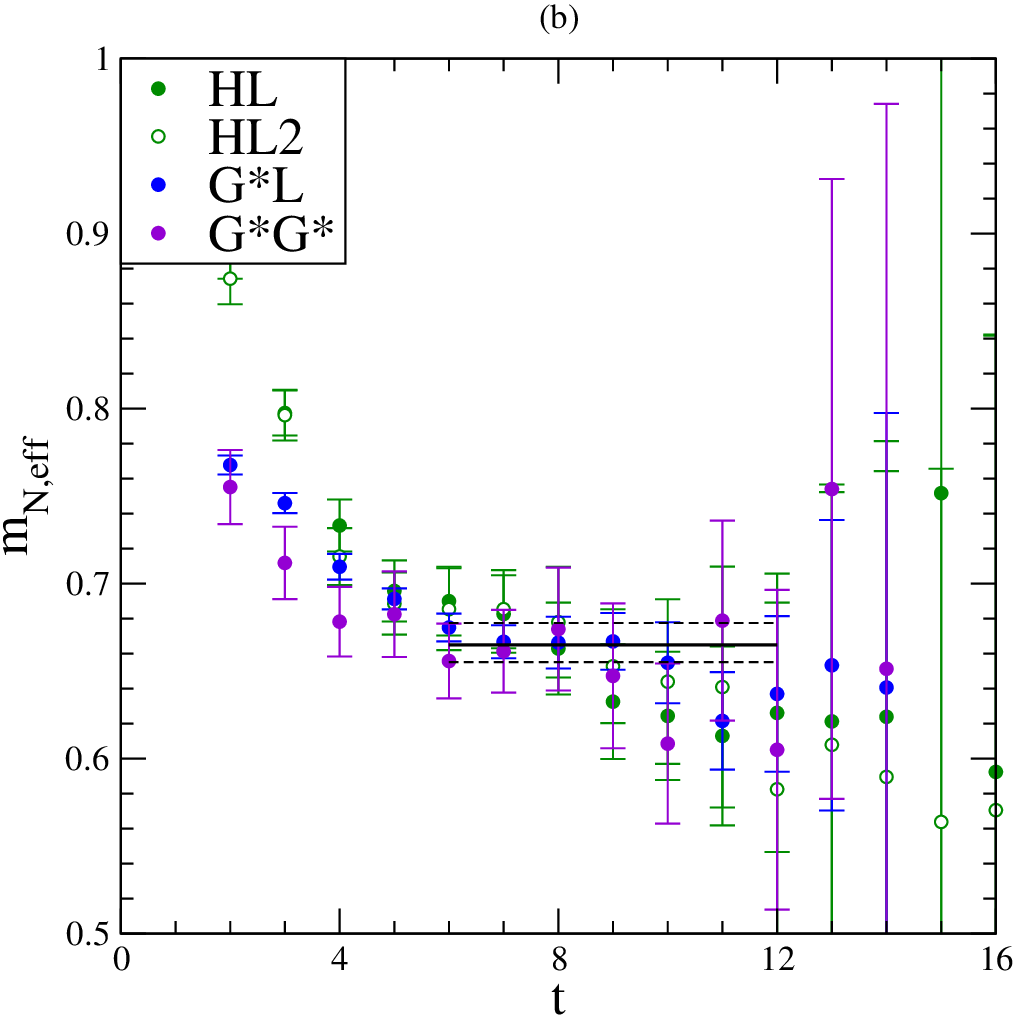}
\caption{\label{fig:n-mass-m005} Effective mass plots for the nucleon
on the $am_u = 0.005$, $V=24^3 \times 64$ ensemble. The different
colours in the effective mass plot correspond to different
smearings. Datasets are labelled with the smearing (i.e. LL). Those
datasets labelled with a 2 use the operator $f_{A_4S,A_4S}(t)$, the
rest use $f_{PS,PS}(t)$. (a) shows the fit before scaling the errors
and (b) shows the fit after rescaling the errors}
\end{figure}

For the case of the nucleon mass on the $am_u = 0.005$, $V=24^3 \times
64$ ensemble (the ensemble with the lightest valence quark mass),
there was some difficulty in judging exactly where the plateau for the
effective mass started. Fitting to different time ranges gave
incompatible results. To account for this, we performed a fit over a
large time range, spanning the multiple potential plateaux. The
incompatibility of the data was reflected in a poor value of $\chi^2$
per degree of freedom (d.o.f) of 4.3. In order to deal with this we
rescaled the errors on all the points in the effective mass plot by
$\sqrt{\chi^2 / \mathrm{d.o.f}}$ and performed a second fit to this
rescaled data. This gave a $\chi^2$/d.o.f of 1, as expected, and a
fitted mass compatible with the best fit value from before, but with a
larger error. The fits to the effective mass on this ensemble before
and after rescaling are shown in Fig. \ref{fig:n-mass-m005}.

The nucleon mass and amplitude are extrapolated linearly to the chiral
limit. The result of the extrapolation for the nucleon mass on the
$24^3 \times 64$ dataset is displayed in Fig.~\ref{fig:n-mass}(c). The
results for the nucleon masses obtained from the fits are summarized
in Table~\ref{tab:results}.

\input{fitrange.tex}

\subsection{Low--energy constants}

\begin{figure}
{\includegraphics[width=.4\textwidth, bb = 0 0 520 490]{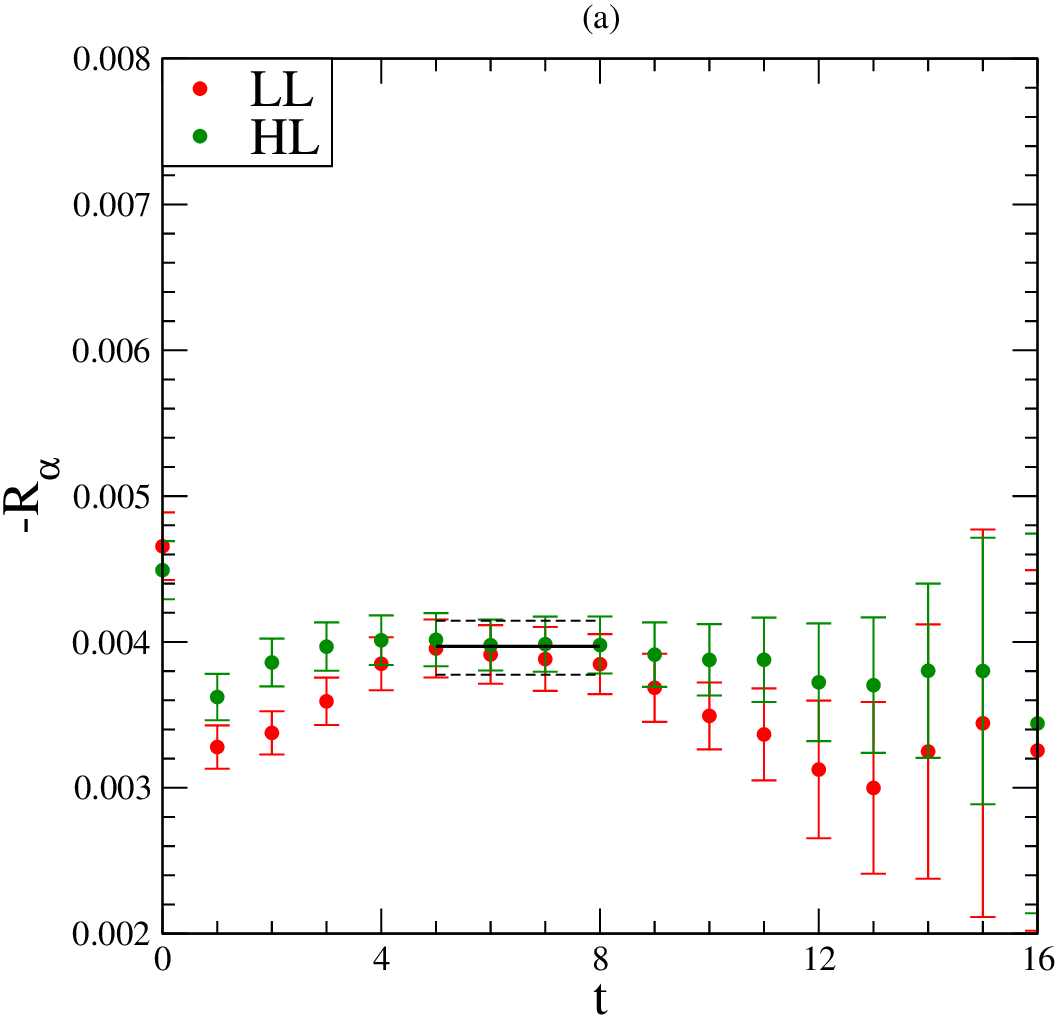}}
{\includegraphics[width=.4\textwidth, bb = 0 0 520 490]{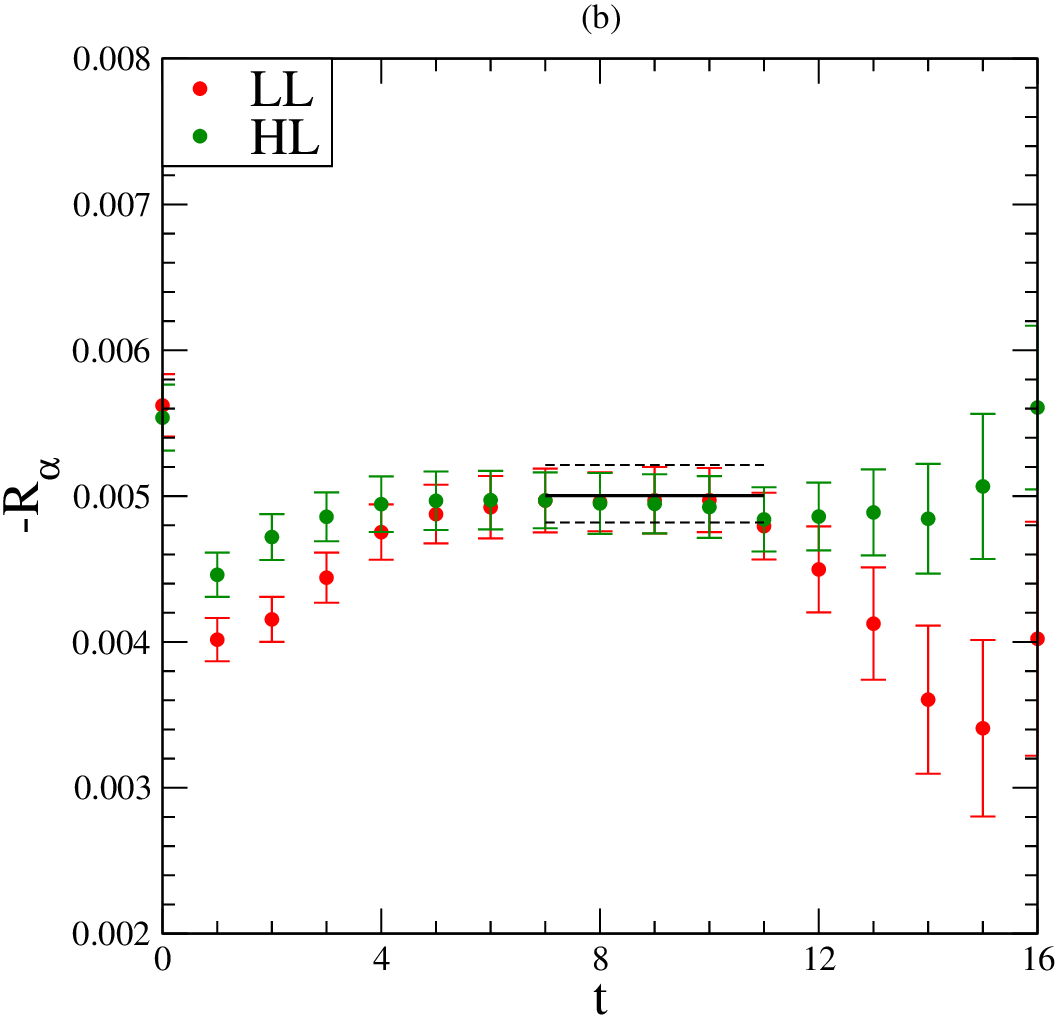}}
{\includegraphics[width=.4\textwidth, bb = 0 0 520 490]{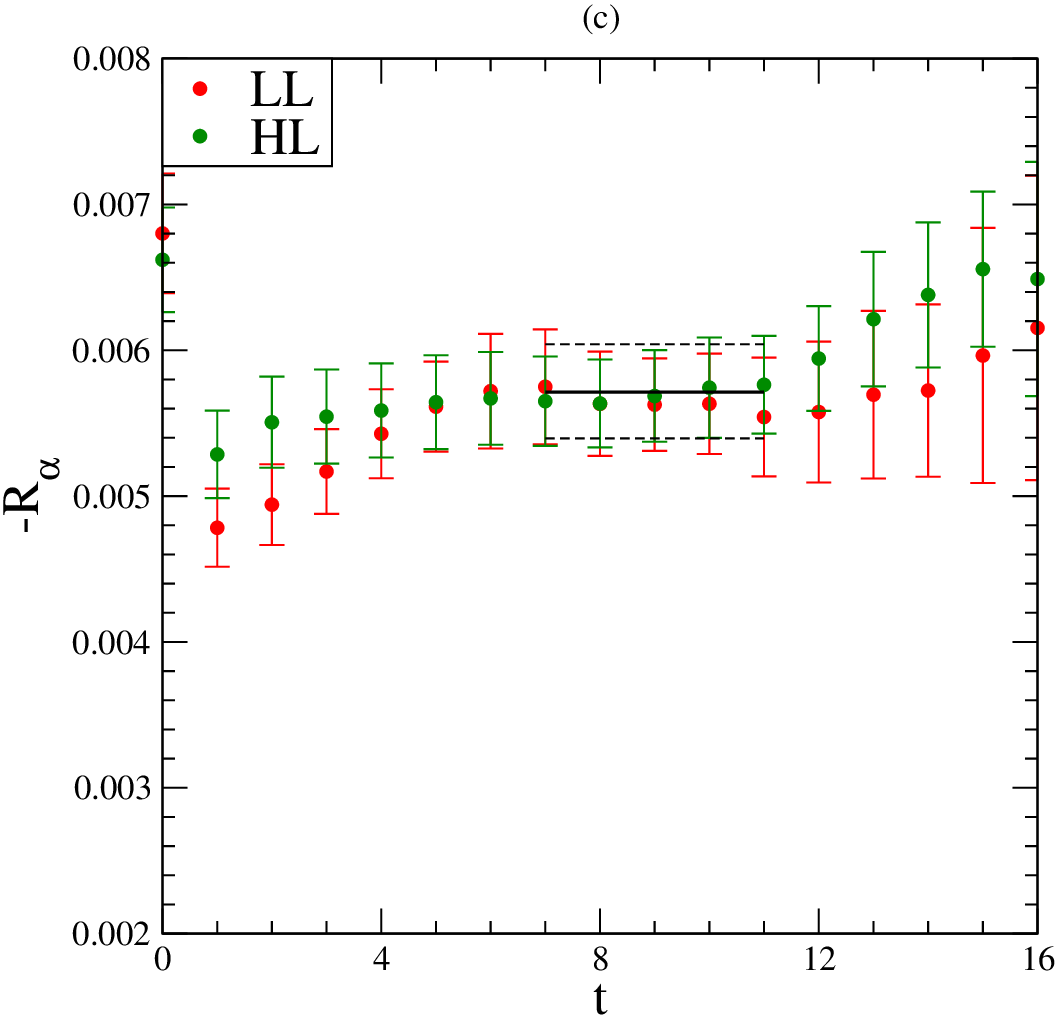}}
{\includegraphics[width=.4\textwidth, bb = 0 0 520 490]{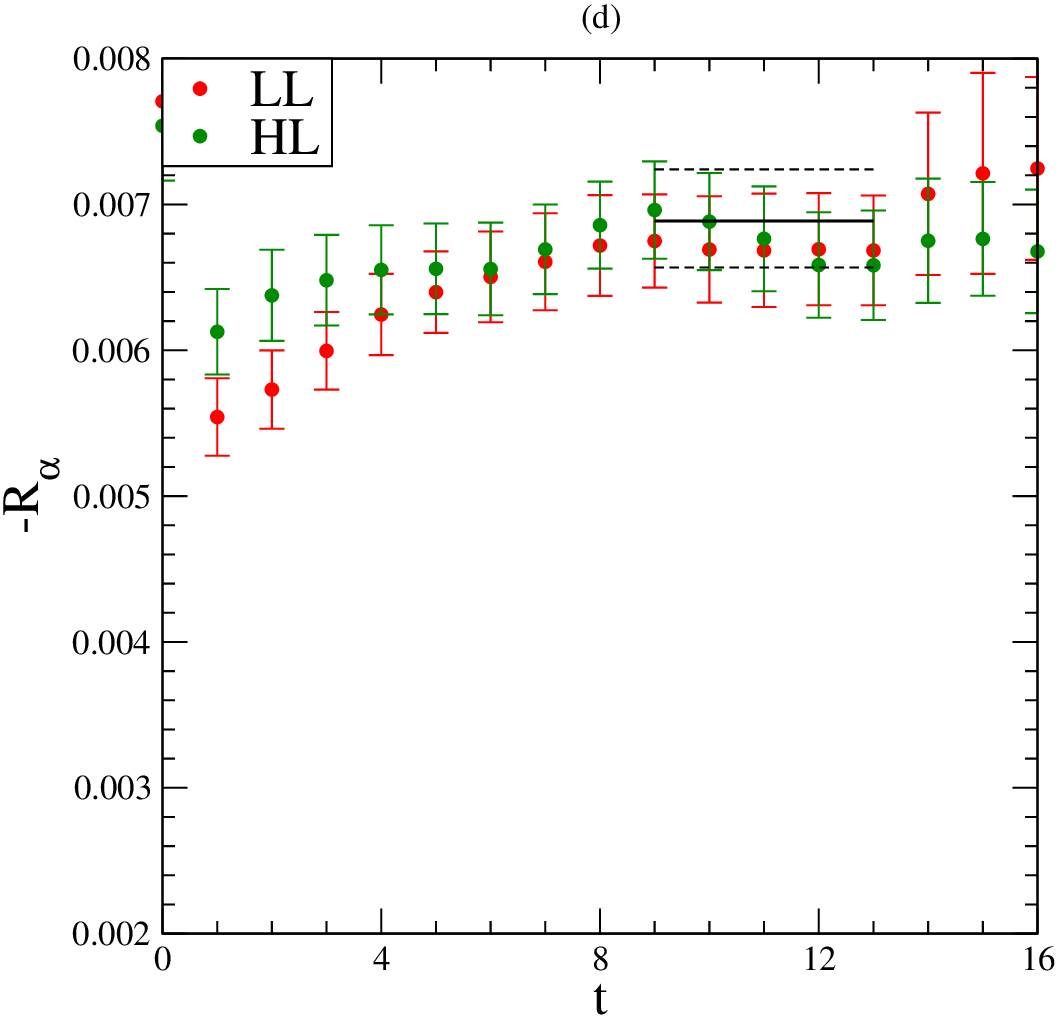}}
\caption{\label{fig:ratio_alpha} The ratio $R_{\alpha}$ in
Eq.~\protect\ref{eq:ratio2pt} for the $24^3 \times 64$ dataset with
$am_u = 0.005, 0.01, 0.02$ and $0.03$ respectively. The different colours
correspond to different source smearing.  Horizontal lines show the
fit to the plateau.}
\end{figure}

As discussed in the previous section, the low--energy parameters
$\alpha$ and $\beta$ appearing in the chiral Lagrangian can be
calculated at leading order through the proton to vacuum matrix
elements:
\begin{align}
  \langle 0 | {\mathcal O}^{RL} | p({\mathbf k},s)\rangle = 
\alpha P_L u({\mathbf k},s), &&
  \langle 0 | {\mathcal O}^{LL} | p({\mathbf k},s)\rangle = 
\beta P_L u({\mathbf k},s),
  \label{eqn:a1}
\end{align}
\begin{align}
  -\langle 0 | {\mathcal O}^{LR} | p({\mathbf k},s)\rangle =  
\alpha P_R u({\mathbf k},s), &&
   -\langle 0 | {\mathcal O}^{RR} | p({\mathbf k},s)\rangle =
\beta P_R u({\mathbf k},s),
  \label{eqn:a2}
\end{align}
where Eq.~\ref{eqn:a2} is obtained from Eq.~\ref{eqn:a1} by parity
transformation. The low--energy constants are obtained from the
asymptotic behaviour of ratios of two--point functions for large
Euclidean time $t$:
\ba
R_{\alpha}(t) & =&2G_N\frac{f_{RL,PS}(t)}{f_{PS,PS}(t)}
\rightarrow \alpha, \\
R_{\beta}(t) & =&2G_N\frac{f_{LL,PS}(t)}{f_{PS,PS}(t)}
\rightarrow \beta.
\label{eq:ratio2pt}
\ea
A typical plateau obtained for $R_\alpha$ is shown in
Fig.~\ref{fig:ratio_alpha}. Two different smearing combinations were
used in the analysis, which correspond respectively to a local and a
smeared interpolating field $\mathcal{O}^{PS}(\vec{x},t)$ for the
nucleon. They both yield consistent results, as shown in the plot. The
values of the low--energy constants were obtained by fitting the data
to a constant, for each value of the quark masses. As for the
spectrum, $\chi^2$ is always defined taking into account the
correlation between different time-slices. The results obtained from
the fits are given in Table~\ref{tab:results}.

\begin{figure}
{\includegraphics[width=.4\textwidth, bb = 0 0 520 490]{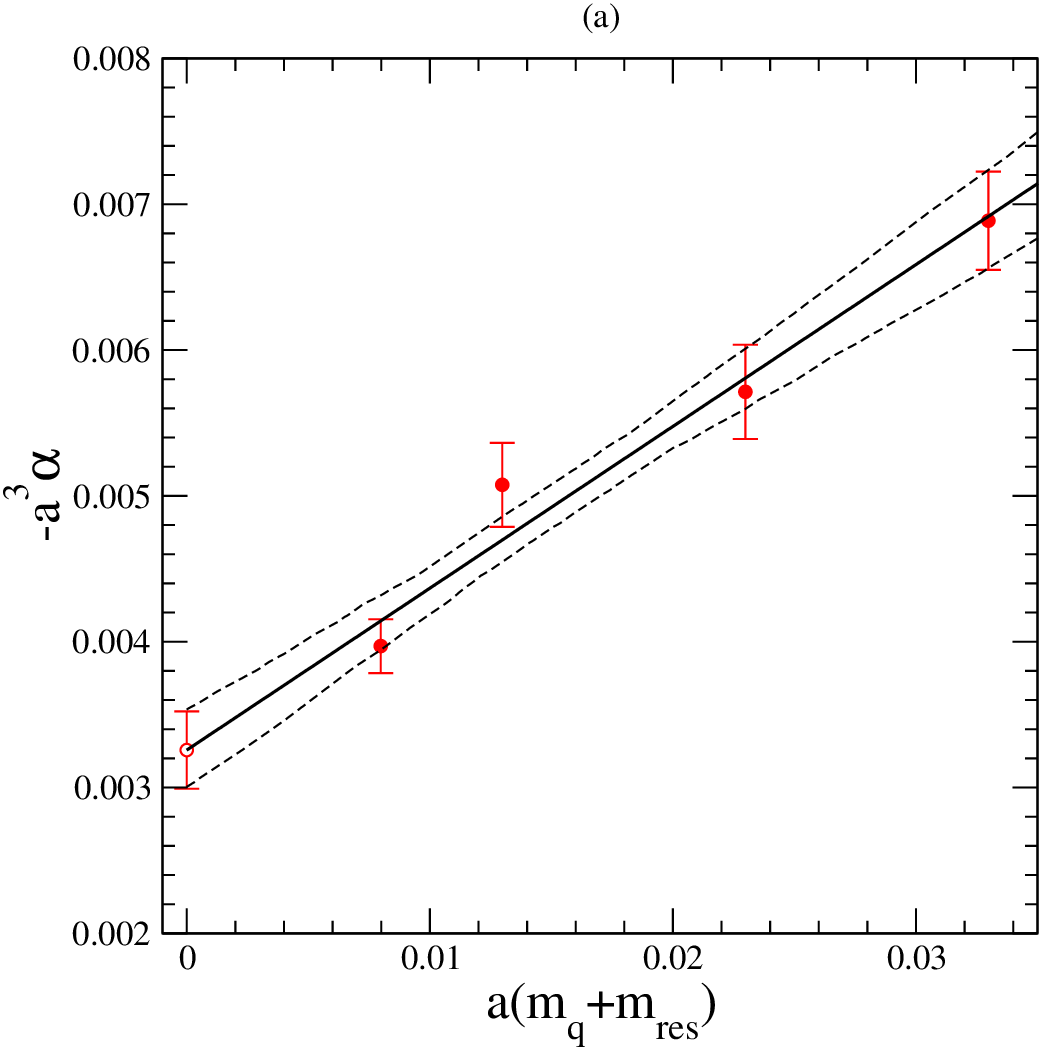}}
{\includegraphics[width=.4\textwidth, bb = 0 0 520 490]{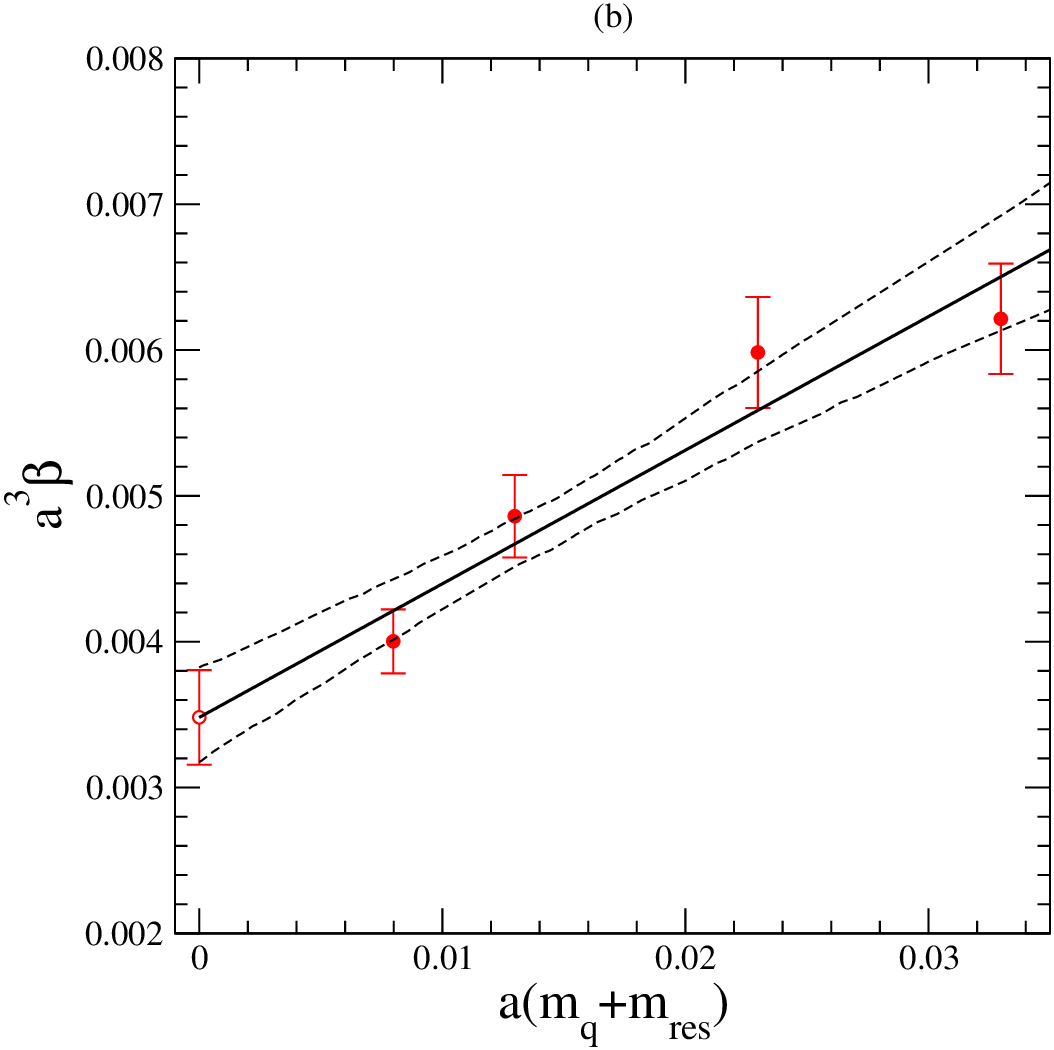}}
\caption{\label{fig:chiralx} Linear chiral extrapolation for the
ratios $R_{\alpha}$ (a) and $R_{\beta}$ (b) for the $24^3\times 64$
dataset.}
\end{figure}

The data points are extrapolated linearly to the chiral limit, as
shown in Fig.~\ref{fig:chiralx}. The data in the mass range studied in
this work appear to be consistent with a linear behaviour, leading to
a good fit for the chiral extrapolation. An uncorrelated $\chi^2$ is
used in this case, since the points at different values of the quark
mass are produced by independent runs.

%
\input{results.tex}

\section{Non--perturbative renormalization}
\label{sec:NPR}

\subsection{RI--MOM mixing matrix}
For the non--perturbative renormalisation of the proton decay matrix
elements we employ the non--perturbative, MOM-scheme, renormalisation
technique of the Rome-Southampton group~\cite{Martinelli:1994ty} as
used by \cite{Aoki:2006ib},\cite{Martinelli:1994ty} and
\cite{Aoki:2007xm}. The application of this technique to proton decay
matrix elements is outlined in~\cite{Aoki:2006ib} which we briefly
summarise.

The operators, $\mathcal{O}^{\Gamma\Gamma^\prime}$, can be classified
according to their symmetry properties under parity ($\mathcal P$) and
the so-called switching transformation ($\mathcal S$). The result of
such classification is summarized in Table~\ref{tab:symmprop}.
\begin{table}[htb]
\begin{center}
\begin{tabular}{c|c|c}
\hline
               & $\mathcal S^-$ & $\mathcal S^+$ \\
\hline
$\mathcal P^-$ & $SS$, $PP$, $AA$     & $VV$, $TT$ \\
\hline
$\mathcal P^+$ & $SP$, $PS$, $-AV$    & $-VA$, $T\tilde{T}$ \\
\hline 
\end{tabular}
\caption{Classification of the $\mathcal{O}^{\Gamma\Gamma^\prime}$
  operators according to their transformation properties under parity
  and switching.\label{tab:symmprop}}
\end{center}
\end{table}
In the presence of chiral symmetry breaking, operators that belong to
the same sector mix under renormalisation. Concentrating on the
$\mathcal S^-$ sectors, the renormalised operators in the parity basis
are defined as:
\begin{equation}
\mathcal{O}^A_{ren} = \tilde Z^{AB}_\mathrm{ND} 
\mathcal{O}^B_{latt}, 
\quad A,B = \{SS,PP,AA\},
\end{equation}
where $A$ and $B$ label the possible choices for
$\Gamma\Gamma^{\prime}$ and $\tilde Z^{AB}_\mathrm{ND}$ is a
$3\times3$ mixing matrix. The same mixing matrix renormalizes the
operators in the sector $\mathcal P^-$ and $\mathcal P^+$. The
chirality basis, which contains the operators of interest for the
nucleon decay matrix elements, consists of
\ba
 LL    &=& \frac{1}{4}(SS+PP) - \frac{1}{4}(SP+PS)\\
 RL    &=& \frac{1}{4}(SS-PP) - \frac{1}{4}(SP-PS)\\
 A(LV) &=& \frac{1}{2}AA - \frac{1}{2}(-AV),
\ea
hence, the mixing matrix in the chirality basis, $Z_{\mathrm{ND}}$,
and in the parity basis are related via:
\begin{equation}
Z_\mathrm{ND} = \mathcal{T} \tilde Z_\mathrm{ND} \mathcal{T}^{-1},
\end{equation}
where
\begin{equation}
\mathcal{T} = \left(
\begin{array}{ccc}
1/4 & 1/4 & 0 \\
1/4 & -1/4 & 0 \\
0 & 0 & 1/2
\end{array}
\right).
\end{equation}
The mixing matrix in the parity basis is computed from the
non--perturbative amputated three--quark vertex function of the
operators in the $\mathcal S^-$ sector as a function of external leg
momentum $p$ after gauge fixing to Landau gauge. The number of
configurations used in the non--perturbative renormalisation is given
in Table~\ref{tab:datasets}. The vertex function is defined as the
amputated Fourier transform of the correlator of $\mathcal{O}^A$ with
three quark spinors:
\begin{equation}
\mathcal G^A(p^2)_{abc\;\alpha\beta\;\gamma\delta} = 
\epsilon^{a^\prime b^\prime c^\prime} (C\Gamma)_{\alpha^\prime\beta^\prime} 
\Gamma^\prime_{\delta\gamma^\prime} \langle
	Q_{\alpha^\prime\alpha}^{a^\prime a}(p)
	Q_{\beta^\prime\beta}^{b^\prime b}(p)
	Q_{\gamma^\prime\gamma}^{c^\prime c}(p) \rangle,
\end{equation}
where
\begin{equation}
Q_{\alpha^\prime \alpha}^{a^\prime a}(p) =  \langle S^{a^{\prime} a^{\prime\prime}}_{\alpha^\prime \alpha^{\prime\prime}}(p)\rangle^{-1} S_{\alpha^{\prime\prime} \alpha}^{a^{\prime\prime} a}(p), 
\quad 
S_{\alpha^\prime \alpha}^{a^\prime a}(p) = \int dx\, e^{-ip.x}\,  S_{\alpha^\prime\alpha}^{a^\prime a}(x),
\end{equation}
$S(x)$ is the quark propagator, and $\Gamma,\Gamma^\prime$ are the
matrices that appear in $\mathcal{O}^A$. 

Introducing the mixing matrix:
\begin{equation}
 M^{AB}(p^2) = \mathcal G^A_{abc\;\alpha\beta\;\gamma\delta} \cdot
  P^B_{abc\;\beta\alpha\;\delta\gamma},
\label{eq:M}
\end{equation}
the renormalisation condition in the RI--MOM scheme reads:
\be\label{eq:ren-condition}
Z_q^{-3/2} \tilde Z_\mathrm{ND}^{BC} M^{CA} = \delta^{BA}
\ee
where $Z_q$ is the quark wavefunction renormalisation; $a,b,c$ are
colour indices and $\alpha$, $\beta$, $\gamma$ and $\delta$ are spin
indices associated with $\Gamma$, $\Gamma^{\prime}$ respectively. The
projection operators,
\begin{eqnarray}
 P^{SS} & = & \frac{1}{96} \epsilon^{abc}(C^{-1})^{\beta\alpha}
  \delta^{\delta\gamma}\\
 P^{PP} & = & \frac{1}{96} \epsilon^{abc}(\gamma_5 C^{-1})^{\beta\alpha}
  \gamma_5^{\delta\gamma}\\
 P^{AA} & = & \frac{1}{384} \epsilon^{abc}(\gamma_5 \gamma_\mu
  C^{-1})^{\beta\alpha} (\gamma_5 \gamma_\mu)^{\delta\gamma},
\end{eqnarray}
are chosen such that the renormalisation condition in
Eq.~\ref{eq:ren-condition} is satisfied in the free field case:
$Z_q=1$, $Z_\mathrm{ND}^{BC} = \delta^{BC}$.

\begin{figure}
{\includegraphics[width=.32\textwidth]{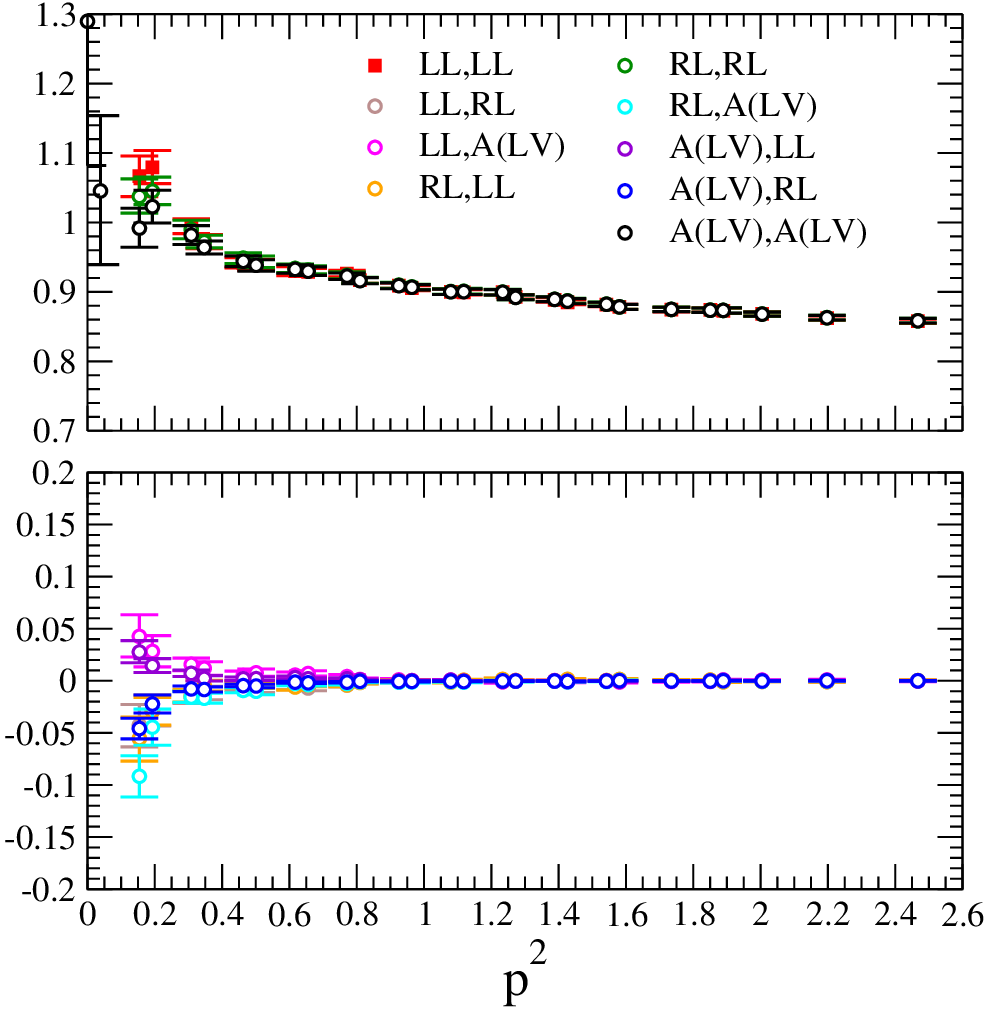}}
{\includegraphics[width=.32\textwidth]{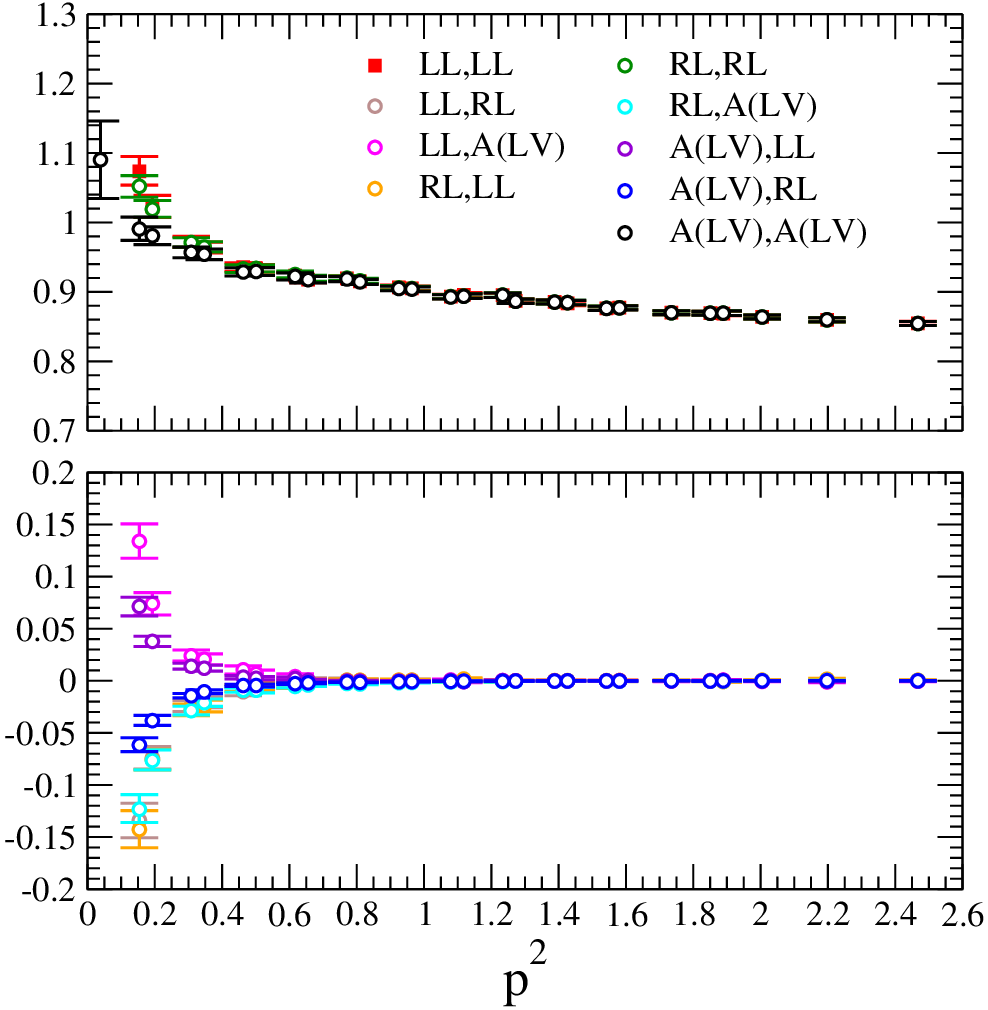}}
{\includegraphics[width=.32\textwidth]{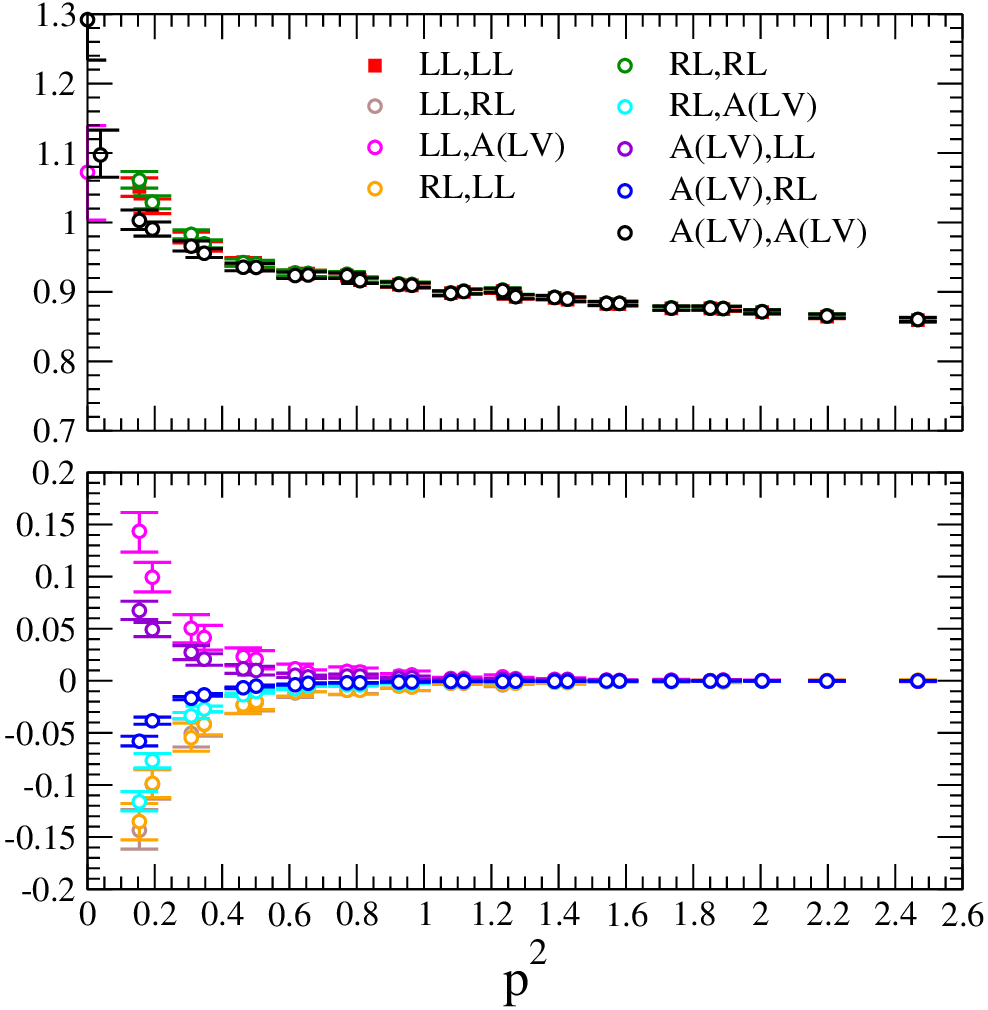}}
\caption{\label{fig:Z_chi} The mixing matrix $M$ in
Eq.~(\protect\ref{eq:M}) in the chirality basis,
$\Gamma\Gamma^{\prime} = \{LL,RL,A(LV)\}$, as a function of the
lattice momentum squared for the $16^3 \times 32$ lattices with $am_u
= 0.01, 0.02$ and $0.03$ (from left to right, respectively). The
off-diagonal mixing between operators is highly suppressed. It is
worthwhile to note that the mass dependence of the mixing matrix is
very mild.}
\end{figure}

Fig.~\ref{fig:Z_chi} shows the mixing matrix, $M^{AB}$,
in the chirality basis as a function of external leg momentum. The set
of momenta used to calculate the mixing matrix is defined by
\ba
p = \left(\frac{2\pi}{L_x}n_x,\frac{2\pi}{L_y}n_y,
\frac{2\pi}{L_z}n_z,\frac{2\pi}{L_t}n_t\right)
\ea
where $L_x=L_y=L_z$ is the spatial size of the lattice and $L_t$ is
the time extent. Combinations of $(n_x,n_y,n_z,n_t)$ such that $-2 \le
n_x,n_y,n_z \le 2$ and $-4 \le n_t \le 4$ are chosen and then averaged
into equal $p^2$ values.

Operator mixing is induced by chiral symmetry breaking. The extent to
which chiral symmetry is broken in the domain wall action is
parameterised by the residual mass, $am_{\rm{res}}$, and the induced
mixing is expected to be suppressed by a factor
$(am_{\mathrm{res}})^2$~\cite{Blum:2001sr}.  It may be seen from
Fig.~\ref{fig:Z_chi} that, in the window of momenta for which
contributions from both hadronic effects (low momenta) and
contributions from discretisation effects (high momenta) are small,
the chiral symmetry afforded by the domain wall fermions suppresses
the mixing between different chirality operators and results in a
mixing matrix which is essentially diagonal. This greatly simplifies
the calculation of the proton decay matrix elements compared to, for
example, Wilson fermions~\cite{Aoki:1999tw}.
%

The matrix $\tilde Z_\mathrm{ND}$ can be obtained from the relation
$M=Z_q^{3/2} \tilde Z_\mathrm{ND}^{-1}$, as shown in
Eq.~\ref{eq:ren-condition}, which requires $Z_q$ to be
computed. Instead, we remove the $Z_q$ dependence, and exploit the
accurate determination of $Z_A = 0.7162(2)$ at the chiral limit, which
was computed from ratios of hadronic matrix elements in
Ref.~\cite{Antonio:2007mh}, together with the average of the amputated
local axial vector and vector bilinear currents, which allows the
evaluation of the factor $\Lambda^A = Z_q/Z_A$. Fig.~\ref{fig:OLL}
shows the average and difference of the amputated local axial vector
and vector bilinear currents. The non-zero difference may be taken as
a measure of the systematic error of the renormalisation constant
arising from the closing of the window where the RI--MOM NPR can be
safely applied. It may be observed that for $(ap)^2 \ge 1.7$ there is
$< 1$\% effect, which is enhanced to $2$\% by extrapolation of $(ap)^2
\rightarrow 0$.

The product $\left(\Lambda^A\right)^{-3/2} M^{-1}$ yields $\tilde
Z_\mathrm{ND}/Z_A^{3/2}$ for each value of the sea quark mass, without
having to deal directly with $Z_q$. At finite lattice spacing, $Z_A$'s
only scale dependence is due to the discretisation error, which starts
at $O(a^2p^2)$. Finally the rotation to the chirality basis and a
linear chiral extrapolation are performed, the latter may be done very
precisely, as the mass dependence is extremely mild, as shown in
Fig.~\ref{fig:Z_chi}. As an example, the $p^2$ dependence of the $LL$
element of the matrix $Z_{\mathrm{ND}}/Z_A^{3/2}$ is displayed in
Fig.~\ref{fig:OLL}.

\subsection{Scheme matching and RG running}

In order to relate the lattice, MOM-scheme, matrix elements at scale
$p$ to those in the $\overline{\mathrm{MS}}$, NDR scheme at some scale $\mu$ we
compute the factor
\begin{equation}
 U^{\overline{\mathrm{MS}}\leftarrow latt}(\mu) = U^{\overline{\mathrm{MS}}}(\mu;p)
  \frac{Z^{\overline{\mathrm{MS}}} (p)}{Z^{MOM}(p)} Z_\mathrm{ND}(p),
  \label{eq:running_factor}	
\end{equation}
where $Z^{\overline{\mathrm{MS}}}(p)/Z^{MOM}(p)$ is the matching factor from
$\overline{\mathrm{MS}}$ scheme to MOM scheme at a scale $p$ calculated using
continuum perturbation theory, and $U^{\overline{\mathrm{MS}}}(\mu;p)$ is the
renormalization group evolution factor from scale $p$ to $\mu$ in the
$\overline{\mathrm{MS}}$ scheme. The matching factor has been computed in
Ref.~\cite{Aoki:2006ib}:
\begin{equation}
 \frac{Z^{\overline{\mathrm{MS}}}}{Z^{MOM}} = 1 + \frac{\alpha_s}{4\pi}
  \left[\frac{433}{180}-\frac{1123}{90}\ln 2
   + \xi \left(\frac{587}{180} - \frac{317}{90}\ln 2\right)\right],
  \label{eq:match}
\end{equation}
where $\xi=0$ as we work in Landau gauge. The $\overline{\mathrm{MS}}$
evolution factor reads
\begin{eqnarray}
  U^{\overline{\mathrm{MS}}}(\mu;p) & = & 
  \left[\frac{\alpha_s(\mu)}{\alpha_s(p)}\right]^{\gamma_0/2\beta_0} 
\left[1+\left(\frac{\gamma_1}{2\beta_0}-\frac{\beta_1\gamma_0}{2\beta_0^2}
          \right)
  \frac{\alpha_s(\mu)-\alpha_s(p)}{4\pi}
  \right],
  \label{eq:running2}\\
 \beta_0 & = & 11-\frac{2}{3}N_f,\; \beta_1 = 102-\frac{38}{3}N_f,\\
  \gamma_0 & = & -4,\;
  \gamma_1 = -\left(\frac{14}{3}+\frac{4}{9}N_f-4\Delta\right),
\end{eqnarray}
where the anomalous dimension of the nucleon decay operator has been
calculated up to two loops in $\overline{\mathrm{MS}}$, NDR
scheme~\cite{Nihei:1994tx} and $\Delta= 0, -10/3$ for $LL, RL$
operators respectively. The value of $\alpha_s(p)$ is obtained by
integrating numerically the four--loop $\beta$ function of
Ref.~\cite{vanRitbergen:1997va}, starting from
$\alpha_s(M_Z)=0.1176(2)$~\cite{PDBook}, and matching the value of
$\alpha_s$ across the $b$, and $c$ thresholds.

\begin{figure}
{\includegraphics[width=.45\textwidth]{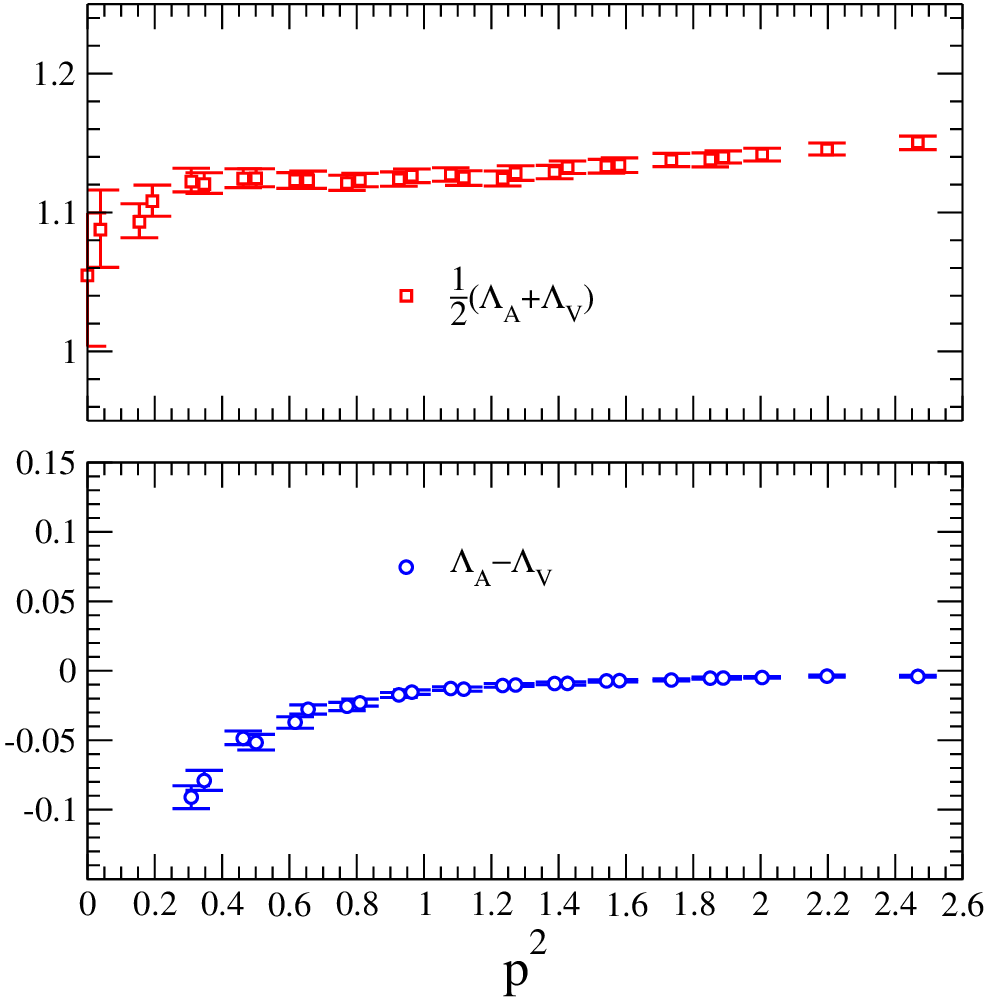}
 \includegraphics[width=.45\textwidth]{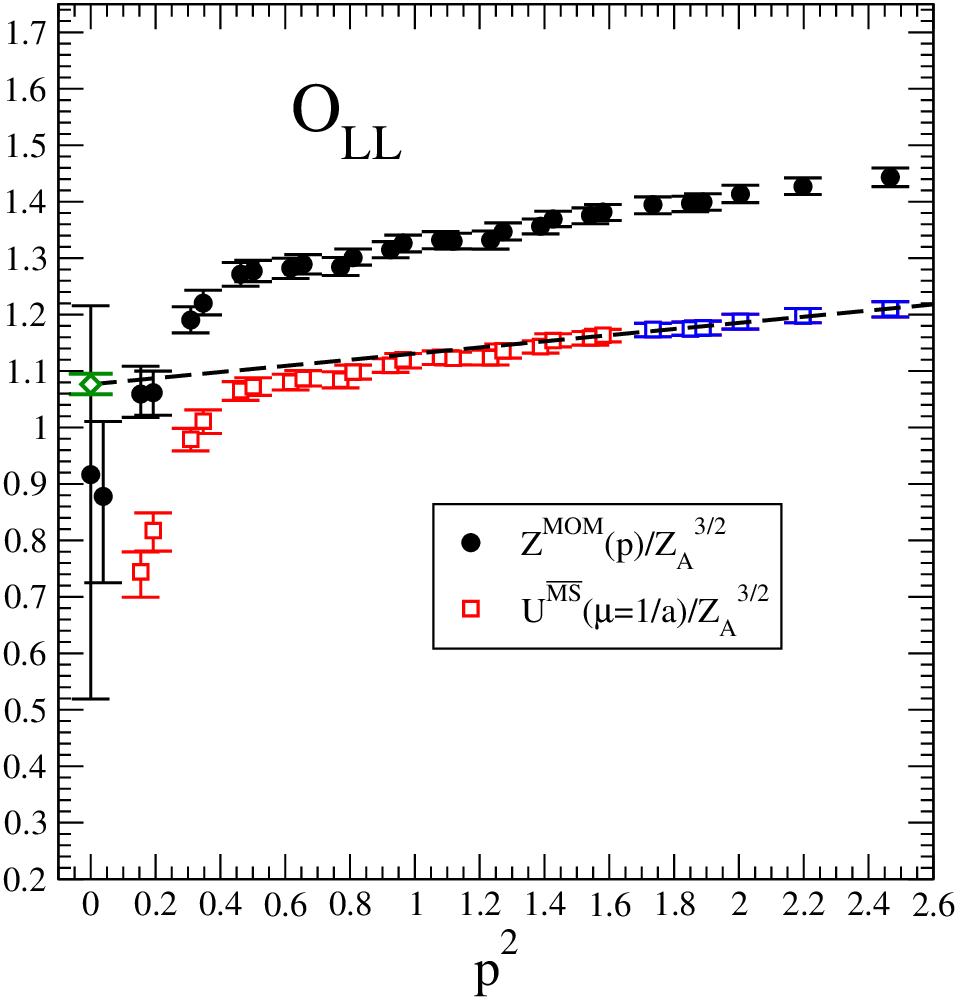}}
\caption{\label{fig:OLL} The figure on the left shows the average and
  the difference of the amputated local axial vector and vector
  bilinear currents as a function of $(ap)^2$.  In the figure on the
  right, the black points show the MOM-scheme renormalisation factor
  in the chiral limit for the $\mathcal{O}^{LL}$ operator normalised
  by the axial current renormalisation factor $Z_A$ as a function of
  the renormalisation scale $(ap)^2$. The red points show the
  renormalization factor in the $\overline{\mathrm{MS}}$ scheme at a scale
  $\mu=1/a$ as a function of the matching scale. The red line shows
  the linear extrapolation in $(ap)^2$, where the blue points are
  those included in the extrapolation.}
\end{figure}

The $\overline{\mathrm{MS}}$ renormalisation factor,
Eq.~\ref{eq:running_factor}, at a fixed scale $\mu = 1/a$ is plotted
in Fig.~\ref{fig:OLL} as a function of the square of the scale at
which the lattice, MOM-scheme, renormalisation calculation was
performed. The remaining momentum dependence, due to $O(a^2p^2)$
discretisation errors, is removed by performing a linear extrapolation
in $(ap)^2$ to $(ap)^2 = 0$, which is also shown in
Fig.~\ref{fig:OLL}. This extraploation is performed over the range
$1.7 < (ap)^2 < 2.5$ where the non-perturbative effect, estimated at
$2$\%, is expected to be small.

Together with the value of $Z_A$ from the hadronic matrix element
ratio and using Eq.~\ref{eq:running2} to run from $\mu = 1/a$ to
$\mu = 2$~GeV we obtain:
\begin{eqnarray}
U^{\overline{\mathrm{MS}}\leftarrow latt}(\mu = 2~\rm{GeV})_{LL} &=& 0.662 \pm 0.010
\nonumber \\ 
U^{\overline{\mathrm{MS}}\leftarrow latt}(\mu = 2~\rm{GeV})_{RL} &=& 0.665 \pm 0.008
\nonumber
\end{eqnarray}
where the error is statistical.

%

\section{Discussion}

\begin{figure}
{ 
  \includegraphics[width=.45\textwidth]{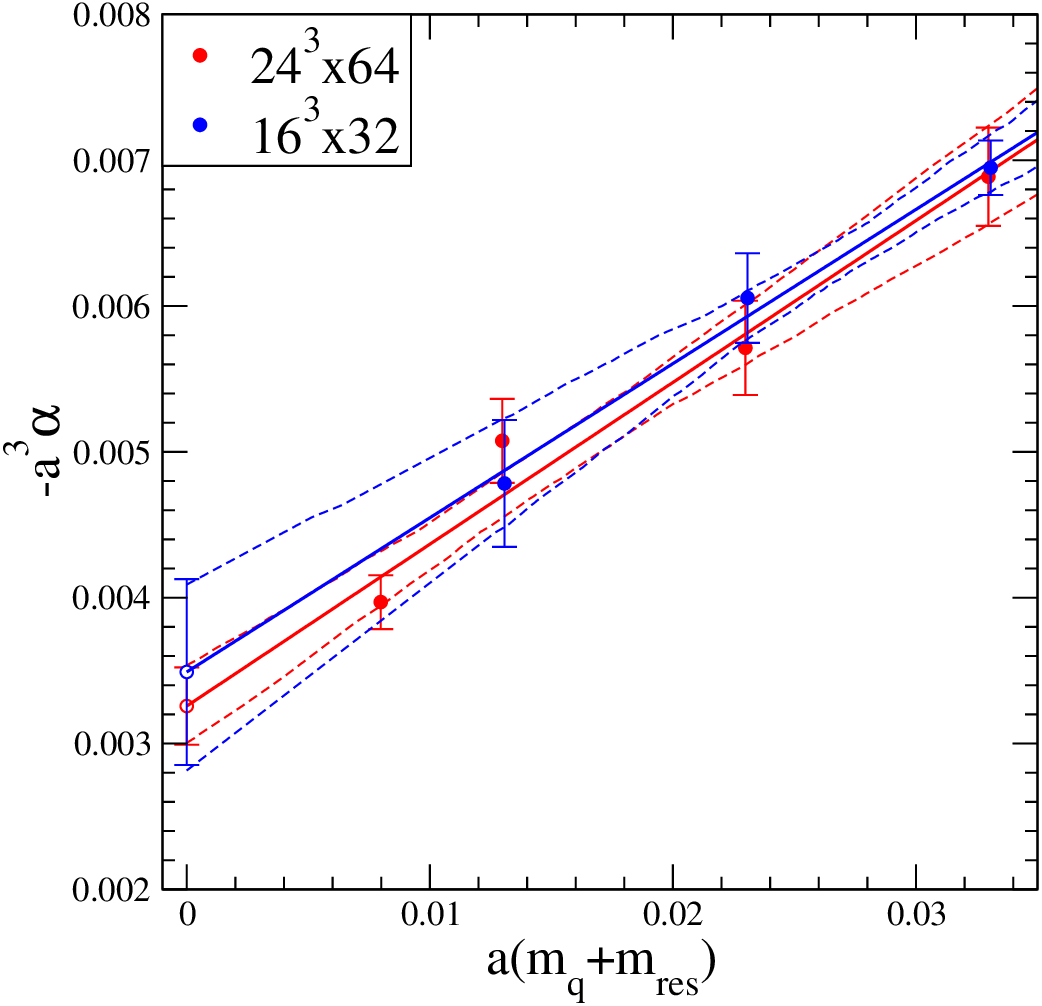}
  \caption{\label{fig:finitesize}The LEC $\alpha$ measured on the two
  different volumes. There are no noticeable finite size effects. The
  chiral extrapolations from the two volumes are shown as white filled
  circles and also agree within errors.}
}
\end{figure}

\begin{figure}
{
  \includegraphics[width=.4\textwidth]{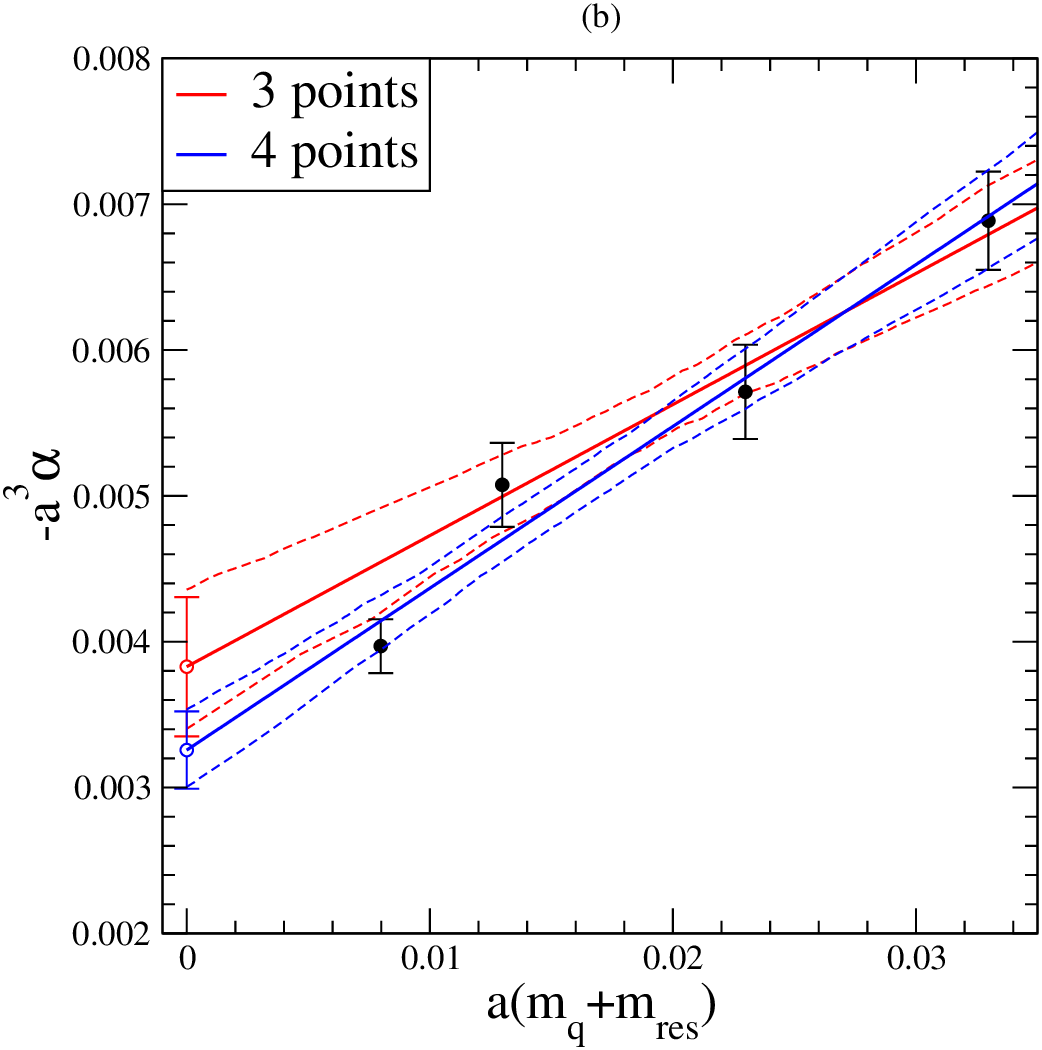}
  \includegraphics[width=.4\textwidth]{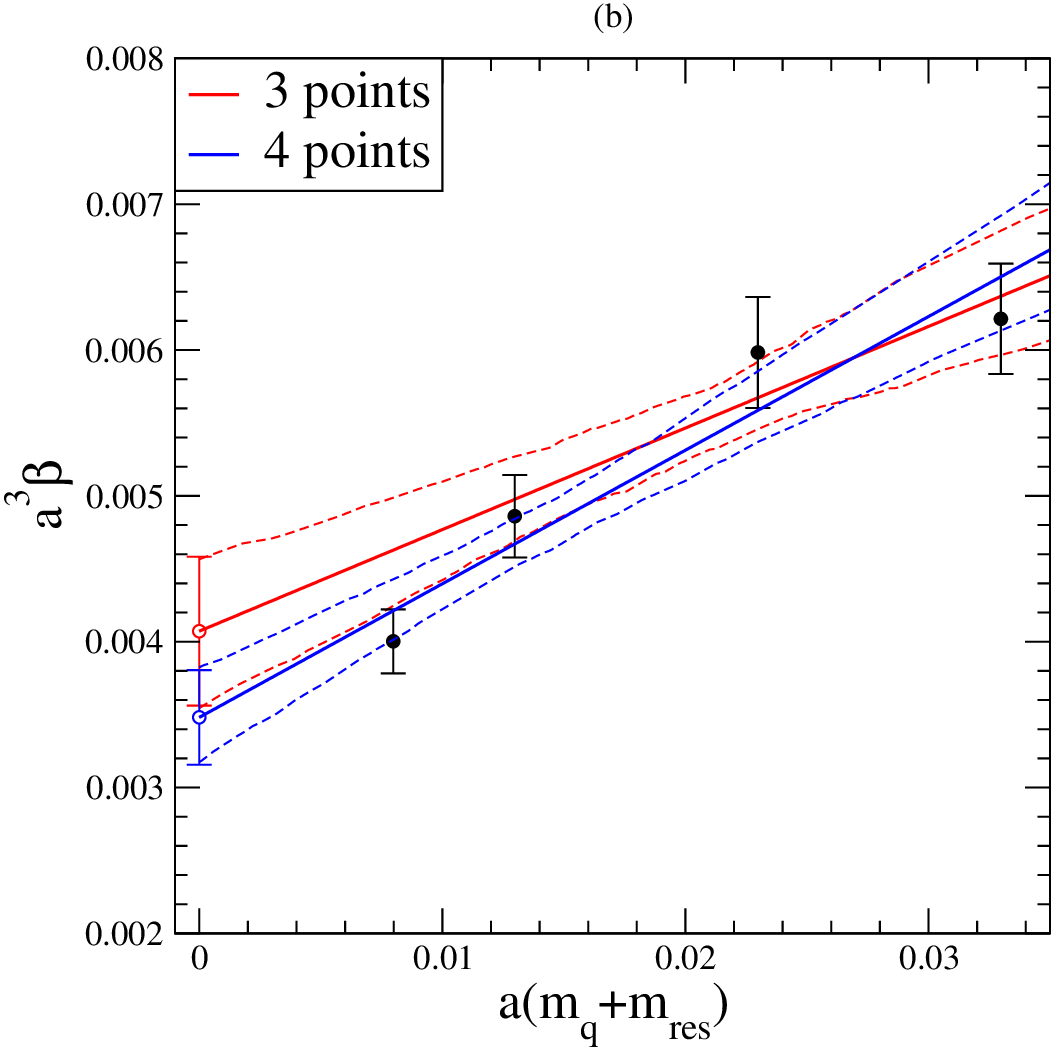}
  \caption{\label{fig:chiralextrap}An extrapolation for $\alpha$ and
$\beta$ both with and without the value from the lightest valence
quark mass point. This gives results differing by 18\% for $\alpha$
and 17\% for $\beta$}. }
\end{figure}

The errors on all quantities so far have been purely statistical.
From the results in Table~\ref{tab:results} we can see that for this
matrix element and for the statistics available, there are no
significant finite volume effects as the results on both volumes agree
within errors.  Fig. \ref{fig:finitesize} shows the agreement for
$\alpha$ between the two volumes.  As discussed in Section
\ref{subsec:nucmassamp}, there is an additional systematic error in
calculating the nucleon mass on the ensemble with the lightest valence
quark mass ($am_u = 0.005$). For a conservative analysis we performed
an extrapolation for $\alpha$ and $\beta$ both with and without this
lightest point. This gave a result which differed by 18\% for $\alpha$
and 17\% for $\beta$ as shown in Fig. \ref{fig:chiralextrap}. We use
this as an estimate of the error in extrapolating to the chiral limit.

\begin{figure}[t]
{\includegraphics[width=.65\textwidth]{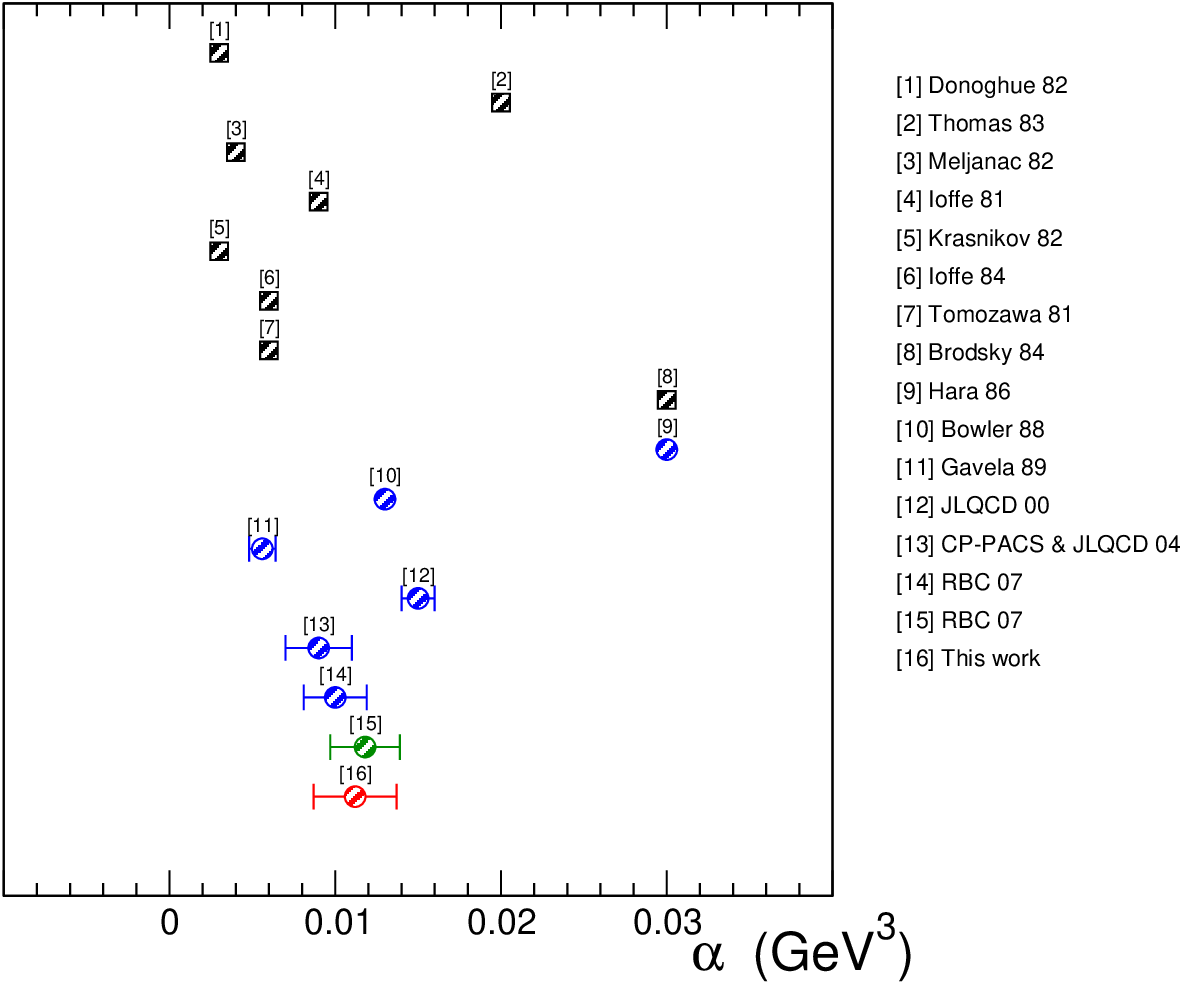}}
\caption{\label{fig:alpha-summary} Summary of computations of the
  hadronic matrix element $\alpha$, as given in
  Tab. \ref{tab:alpha-summary}.  Square points correspond to QCD model
  calculations, blue circles correspond to $N_f = 0$ lattice QCD
  calculations, the green circle is from $N_f = 2$ and the result from
  our $N_f = 2+1$ calculation is shown in red. }
\end{figure}

It should be noted that in our simulation, the strange quark mass is
held fixed and hence in the extrapolation, only the light quarks are
taken to the chiral limit.  However, if we compare our result with the
$N_f = 2$ result from \cite{Aoki:2006ib} we see there is very good
agreement (see Fig. \ref{fig:alpha-summary}). For $N_f = 2$, the
strange quark mass is effectively infinite, the agreement signifies
that $\alpha$ and $\beta$ have little dependence on the strange sea
quark mass.

For the NPR, we estimate a systematic error of $8$\% which is
dominated by the error from truncating the perturbative expansion for
the matching factor at order $\alpha_s^2$ in
Eq. \ref{eq:match}.

Adding all of these uncertainties in quadrature, and together with the values for the matrix
elements in Table~\ref{tab:results}, we estimate the low--energy parameters renormalised at
$\mu = 2$~GeV to be:

\ba
\alpha &=& -0.0112 \pm 0.0012_{(\mathrm{stat})} \pm 0.0022_{(\mathrm{syst})}~\mathrm{GeV}^3
\label{eq:finalalpha}\\
\beta  &=&  0.0120 \pm 0.0013_{(\mathrm{stat})} \pm 0.0023_{(\mathrm{syst})}~\mathrm{GeV}^3.
\label{eq:finalbeta}
\ea

The results for various determinations of $\alpha$ are summarized in
Fig.~\ref{fig:alpha-summary}. The agreement between recent lattice
computations suggests that lattice QCD is being successful at
determining the low-energy constants describing nucleon decay with
increasingly smaller systematic uncertainty.

The indirect computation of the proton lifetime has a further
non-linear systematic error, due to the use of chiral perturbation
theory in a kinematic regime where the pion has a large momentum. The
relevant matrix element has been computed using both the indirect and
direct methods in Ref. \cite{Aoki:2006ib} where sizeable differences
were seen between the two methods. For the case of the matrix elements
in Eq. \ref{eq:me1} and \ref{eq:me2}, the
indirect method was found to give an estimate for the matrix element
of about two times larger than the direct method.

\input{alpha-summary.tex}

Finally, let us discuss one way to use our result to discriminate
between GUTs. The proton partial decay width in a generic channel can
be split into
\be \Gamma = \mathrm{LEC}^2 \times A_{\mathrm{QCD}}
\times A_{\mathrm{GUT}},
\ee
where LEC is the low--energy constant,
$\alpha$ or $\beta$, that we calculate earlier in this paper,
$A_{\mathrm{QCD}}$ contains information from QCD parameters and
$A_{\mathrm{GUT}}$ contains all the information about the underlying
high--energy theory, including constants from the GUT. For the Minimal
SUSY SU(5) GUT, expressions for the lifetime have been calculated for
several decay modes in Ref. \cite{Hisano:1992jj}.

Dimensional Analysis gives the value of the proton lifetime as
$\Gamma_N \sim \alpha_{\mathrm{GUT}}^2m_p^5 /
M_{\mathrm{GUT}}^4$. Taking $M_{\mathrm{GUT}} \approx 10^{16}$
\cite{PDBook} and obtaining $\alpha_{\mathrm{GUT}}$ by running the
strong coupling up to the GUT scale gives $\Gamma_N \sim 10^{-68}$
GeV. The natural scale for $A_{\mathrm{GUT}}$ is
$M_{\mathrm{GUT}}^{-4}$.  The values of the low--energy constant that
we have computed, together with the values of the quantities in
$A_{\mathrm{QCD}}$ \cite{PDBook} and the experimental bounds on the
proton life time \cite{Nishino:2008,Kobayashi:2005pe}, allow us to put
bounds on $A_{\mathrm{GUT}}$ for different decay modes as summarized
in Table \ref{tab:lifetimes}. The bounds quoted are at a 68\%
confidence level.  The different decay modes provide more or less
stringent bounds on $A_{\mathrm{GUT}}$.  The viability of any GUT can
be assessed by checking wether the relevant bounds are satisfied.

\input{lifetimes.tex}

As a simple example, for the decay mode $p\rightarrow e^+\pi^0$ via X
boson exchange in the minimal SU(5) SUSY GUT,
$A_{\mathrm{GUT}}(p\rightarrow e^+\pi^0)$ is given by
\cite{Hisano:2000dg} to be \be A_{\mathrm{GUT}}(p\rightarrow e^+\pi^0)
= \frac{g_5^4A_R^2}{M_X^4} \left|1+\left(1+|V_{ud}|^2\right)^2\right|
\ee where $g_5$ is the unified coupling at the GUT scale, $M_X$ is the
mass of the X boson $\approx M_{GUT}$, $A_R$ is the renormalization
factor and $V_{ud}$ a CKM matrix element. Using the value of $A_R$
given in \cite{Hisano:2000dg}, we can put a bound on the $X$ boson
mass of $M_X > 5\times10^{15}\mathrm{GeV}$.

The decay widths of the channels involving colour triplet Higgs
exchange can be calculated and involve the LEC $\beta$ (see
Ref. \cite{Hisano:1992jj}). The analysis in
Ref. \cite{Murayama:2001ur} uses a conservative choice of 
$\beta = 0.003 \mathrm{GeV}^3$ at a scale of 1 GeV to constrain the mass of the
colour triplet Higgs sufficiently to rule out the minimal SUSY SU(5)
GUT. The higher value calculated in this work (running our value of
$\beta$ to a scale of 1 GeV gives $\beta = 0.0109\pm23~\mathrm{GeV}^3$
if we use Eq. \ref{eq:running2}) gives an even stronger constaint on
the mass of the colour triplet Higgs and so confirms the fact that the
minimal SUSY SU(5) GUT has been ruled out.

The uncertainty on $\alpha^2$ is $45\%$ and on $\beta^2$ is
 $43\%$. These are higher than the uncertainties on the factors
 $A_{\mathrm{QCD}}$ which for all channels is $\approx 8\%$.  A
 factor of $\approx 2$ reduction in the uncertainty of the LECs would make them
 comparable with the uncertainties of $A_{\mathrm{QCD}}$, which is
 dominated by the uncertainties of $D$, $F$ and $f_\pi$. As $M_X \sim
 \sqrt\alpha$, an error of $45\%$ on $\alpha^2$ corresponds to an
 error of $11\%$ on the bound for $M_X$. Reducing the uncertainty on
 $\alpha$ by a factor of two would therefore reduce the uncertainty
 from $\alpha$ on the bound for $M_X$ to $6\%$.

\section*{Acknowledgments}

The calculations reported here were done on the QCDOC computers
\cite{Boyle:2005qc,Boyle:2003mj,Boyle:2005fb} at Columbia University,
Edinburgh University, and at Brookhaven National Laboratory (BNL).
At BNL, the QCDOC computers of the RIKEN-BNL Research Center and
the USQCD Collaboration were used.  The software used includes:  the
CPS QCD codes {\tt http://qcdoc.phys.columbia.edu/chulwoo\_index.html},
supported in part by the USDOE SciDAC program; the
BAGEL {\tt http://www.ph.ed.ac.uk/\~{}paboyle/bagel/Bagel.html} assembler
kernel generator for many of the high-performance optimized kernels;
and the UKHadron codes.

AS (BNL) was partially supported by the U.S.\ DOE under
contract DE-AC02-98CH10886. The work of the Edinburgh authors was
supported by PPARC grants PP/D000238/1 and PP/C503154/1. The former
directly supported CMM. PAB acknowledges support from RCUK and
LDD is funded through an STFC advanced fellowship. The Edinburgh QCDOC
system was funded by PPARC JIF grant PPA/J/S/1998/00756 and operated
through support from the Universities of Edinburgh, Southampton and
Wales Swansea, and from STFC grant PP/E006965/1.-

Computations for this work were carried out in part on facilities
of the USQCD Collaboration, which are funded by the Office of Science
of the U.S. Department of Energy.  We thank RIKEN, BNL and the U.S.\
DOE for providing the facilities essential for the completion of
this work.

We thank Masato Shiozawa for communicating the latest Super-Kamiokande
results.

\end{document}

%% file: user_def.tex
\newcommand{\ba}{\begin{eqnarray}}
\newcommand{\ea}{\end{eqnarray}}
\newcommand{\bas}{\begin{eqnarray*}}
\newcommand{\eas}{\end{eqnarray*}}
\newcommand{\be}{\begin{equation}}
\newcommand{\ee}{\end{equation}}
\newcommand{\bes}{\begin{equation*}}
\newcommand{\ees}{\end{equation*}}
\newcommand{\bi}{\begin{itemize}}
\newcommand{\ei}{\end{itemize}}
\newcommand{\lf}{\left}
\newcommand{\rr}{\right}
\newcommand{\nn}{\nonumber}

\newcommand{\bcentre}{\begin{center}}
\newcommand{\ecentre}{\end{center}}
 
\newcommand{\Tr}{\mathrm{Tr}}

\newcommand{\slsh}[1]{/\!\!\!#1}

\font\tenmsb=msbm10 scaled\magstep1
\font\sevenmsb=msbm7 scaled\magstep1
\font\fivemsb=msbm5 scaled\magstep1
\newfam\msbfam
\textfont\msbfam=\tenmsb
\scriptfont\msbfam=\sevenmsb
\scriptscriptfont\msbfam=\fivemsb

\newcommand{\lslash}[1]{#1\!\!\!\!\;\slash}

\newcommand{\bra}[1]{\langle#1|}
\newcommand{\ket}[1]{|#1\rangle}

\def\ln{\mathop{\rm ln}}		    

\def\su3{SU(3)}

\def\tD{\mbox{D}\kern-0.65em\raise0.15ex\hbox{/}\kern0.15em} 
\def\sD{\mbox{\scriptsize D}\kern-0.5em\raise0.15ex\hbox{\scriptsize/}}
\def\ssD{\mbox{\tiny D}\kern-0.42em\raise0.15ex\hbox{\tiny/}}

\def\dslash{\hbox{\(\partial\)}\kern-0.5em\raise0.15ex\hbox{/}} 

\def\su3{SU(3)}



%% file: datasets.tex
\begin{table}[htb]
\begin{center}
\begin{tabular}{cc|ccccc|ccccc}
\hline
&& \multicolumn{5}{c|}{Matrix Elements} & \multicolumn{5}{c}{NPR}\\
$V\times L_s$&$am_{ud}$&$N_{\rm{traj}}$   & $\Delta$&$N_{\rm{cfg}}$&$N_{\rm{src}}$&$N_{\rm{bin}}$&$N_{\rm{traj}}$&$\Delta$&$N_{\rm{cfg}}$&$N_{\rm{src}}$&$N_{\rm{bin}}$\\
\hline
\hline
                       & 0.01  & 500-4000   & 10   & 175          & 4           & 8             & 1000-4000     & 40     & 75          &      4       &  1\\
$16^3\times32\times16$ & 0.02  & 500-4000   & 10   & 175          & 4           & 8             & 1000-4000     & 40     & 75          &      4       &  1\\
                       & 0.03  & 500-7580   & 10   & 177          & 2           & 8             & 1000-4000     & 40     & 75          &      4       &  1\\
\hline
                       & 0.005 & 900-4500   & 10   & 90           & 2           & 8             &               &        &              &             &  \\
                       &            &            & 10   & 90           & $2\times2$  & 8             &               &        &              &             &  \\
$24^3\times64\times16$ & 0.01  & 1500-3860  & 10   & 59           & 2           & 8             & & & & &\\
                       &            &            & 40   & 59           & 2           & 2             &               &        &              &             &  \\
                       & 0.02  & 1800-3600  & 10   & 45           & 2           & 8             & & & & &\\
                       &            &            & 40   & 45           & 2           & 2             &               &        &              &             &  \\
                       & 0.03  & 1020-3060  & 20   & 51           & 1           & 2             & & & & &\\
                       &            &            & 40   & 51           & 1           & 1             &               &        &              &             &  \\
\cline{1-7}
\end{tabular}
\end{center}
\caption{RHMC 2+1 flavour datasets used for the non-perturbative renormalization and matrix element calculation. $V$ is the space-time volume of the lattice, $L_s$ is the extent of the fifth dimension, $am_{ud}$ is the up sea quark mass (the strange sea quark mass is kept fixed at 0.04), $N_{\rm{traj}}$ is the lowest to highest trajectories analysed with matrix elements calculated every $\Delta$ trajectories, $N_{\rm{cfg}}$ is the number of configurations, $N_{\rm{src}}$ is the number of quark propagators solved with different source locations and $N_{\rm{bin}}$ is the bin size. $24^3\times64\times16$ data were generated for the non-perturbative renormalization calculation, however, it was only used as a check for finite volume errors and so does not appear here. For the $24^3\times64\times16$ matrix element data, there are two independent runs for each of the sea quark masses. These independent runs used different smearings, $\Delta$, source locations and $N_{\rm{src}}$.}
\label{tab:datasets}
\end{table}

%% file: fitrange.tex
\begin{table}[htb]
\begin{center}
\begin{tabular}{c||cccccc||cccccc}
\hline
      &\multicolumn{6}{c||}{$V=16^3\times32$}                              &\multicolumn{6}{c}{$V=24^3\times64$}\\
$m_u$ &         &              & \multicolumn{4}{c||}{Fit Range}           &      &                 &  \multicolumn{4}{c}{Fit Range}\\
      & Smearing& $O^{\Gamma\Gamma}$ & $m_N$ & $G_N$ & $\alpha$ & $\beta$  & Smearing &  $O^{\Gamma\Gamma}$ & $m_N$ & $G_N$ & $\alpha$ & $\beta$\\
\hline
0.005 &         &              &       &       &          &                & LL   & $O^{PS}$         & -     & 9-12  & 5-8      & 5-9        \\
      &         &              &       &       &          &                & HL   & $O^{PS}$         & 6-12  & -     & 4-10     & 4-9        \\
      &         &              &       &       &          &                & HL   & $O^{A_4S}$       & 6-12  & -     & -        & -       \\ 
      &         &              &       &       &          &                & G*L  & $O^{PS}$         & 6-12  & -     & -        & -        \\
      &         &              &       &       &          &                & G*G* & $O^{PS}$         & 6-12  & -     & -        & -        \\
\hline
0.01  &LL   & $O^{PS}$         & 9-12  & 9-12  & 5-8      & 5-8            & LL   & $O^{PS}$         & 9-12  & 9-12  & 7-11     & 5-10       \\    
      &LL   & $O^{A_4S}$       & 9-12  & -     & -        & -              & LL   & $O^{A_4S}$       & 9-12  & -     & -        & -        \\ 
      &GL   & $O^{PS}$         & 9-12  & -     & 3-8      & 3-8            & GL   & $O^{PS}$         & 8-12  & -     & 7-11     & 4-10        \\
      &GL   & $O^{A_4S}$       & 9-12  & -     & -        & -              & GL   & $O^{A_4S}$       & 8-12  & -     & -        & -        \\
      &         &              &       &       &          &                & G*L  & $O^{PS}$         & 7-12  & -     & -        & -       \\
\hline
0.02  &LL   & $O^{PS}$         & 9-12  & 9-12  & 6-12     & 5-12           & LL   & $O^{PS}$         & 9-10  & 9-12  & 7-11     & 5-10       \\
      &LL   & $O^{A_4S}$       & 9-12  & -     & -        & -              & LL   & $O^{A_4S}$       & 9-10  & -     & -        & -      \\ 
      &GL   & $O^{PS}$         & 8-12  & -     & 6-10     & 4-14           & HL   & $O^{PS}$         & 9-11  & -     & 7-11     & 7-10       \\
      &GL   & $O^{A_4S}$       & 8-12  & -     & -        & -              & HL   & $O^{A_4S}$       & 9-11  & -     & -        & -      \\
      &GG   & $O^{PS}$         & 8-11  & -     & -        & -              & G*L  & $O^{PS}$         & 9-11  & -     & -        & -      \\
      &GG   & $O^{A_4S}$       & 8-11  & -     & -        & -              &      &                  &       &       &          &           \\    
\hline
0.03  &LL   & $O^{PS}$         & 10-12  & 9-12  & 7-11     & 6-11           & LL   & $O^{PS}$         & 10-12  & 9-12  & 9-13     & 9-11      \\
      &LL   & $O^{A_4S}$       & 10-12  & -     & -        & -              & LL   & $O^{A_4S}$       & 10-12  & -     & -        & -      \\ 
      &GL   & $O^{PS}$         & 8-12   & -     & 7-11     & 8-12           & HL   & $O^{PS}$         & 9-12  & -     & 9-13     & 9-11     \\
      &GL   & $O^{A_4S}$       & 8-12   & -     & -        & -              & HL   & $O^{A_4S}$       & 9-12  & -     & -        & -     \\
      &GG   & $O^{PS}$         & 8-12   & -     & -        & -              & G*L  & $O^{PS}$         & 8-12  & -     & -        & -     \\
      &GG   & $O^{A_4S}$       & 8-12   & -     & -        & -              &      &                  &       &       &          &  \\            
\hline
\end{tabular}
\caption{Smearings, operators and fit ranges used for the calculation
of nucleon masses, nucleon amplitudes and matrix elements.\label{tab:fitranges}}
\end{center}
\end{table}

%% file: results.tex
\begin{table}[htb]
\begin{center}
\begin{tabular}{cc|ccc}
\hline
$V\times L_s$          & $am_{ud}/am_{s}$  & $am_{N}$  & $a^3\alpha$   & $a^3\beta$\\
\hline
\hline
                       & 0.03/0.04         & 0.908(6)  & -0.00695(19) & $0.00719(21)$\\
$16^3\times32\times16$ & 0.02/0.04         & 0.819(8)  & -0.00605(31) & $0.00606(30)$\\
                       & 0.01/0.04         & 0.722(19) & -0.00478(43) & $0.00511(47)$\\
                       & chiral            &           & -0.00349(64) & $0.00369(63)$\\
\hline
                       & 0.03/0.04         & 0.892(10) & -0.00689(33) & $0.00621(38)$\\
$24^3\times64\times16$ & 0.02/0.04         & 0.805(12) & -0.00571(32) & $0.00598(38)$\\
                       & 0.01/0.04         & 0.720(10) & -0.00508(29) & $0.00486(28)$\\
                       & 0.005/0.04        & 0.671(5)  & -0.00397(18) & $0.00400(22)$\\
                       & chiral            &           & -0.00326(27) & $0.00348(32)$\\
\hline
\end{tabular}
\end{center}
\caption{Results from fits described in this paper. The nucleon masses the LECs
  $\alpha$ and $\beta$ are reported as a function of the quark masses, for both lattices used in this study.                                                                      The results of linear chiral extrapolations are also reported in the last line of each column. All the results
  are given in units of the lattice spacing $a\approx0.12~\mathrm{fm}$.}
\label{tab:results}
\end{table}

%% file: alpha-summary.tex
\begin{table}
 \label{tab:alpha-summary}
\begin{tabular}{clccl}
 \hline
 \hline
 & & $|\alpha|$ [GeV$^3$] & $|\beta|$ [GeV$^3$] & \\
 \hline
 \hline
 & Donoghue and Goldwich \cite{Donoghue:1982jm} & 0.003 & & Bag model\\
 & Thomas and McKellar \cite{Thomas:1983ch} & 0.02 & & Bag model\\
 & Meljanac et al. \cite{Meljanac:1981xd} & 
 0.004 & & Bag model\\
 QCD model & Ioffe \cite{Ioffe:1981kw} & 0.009 &  & Sum rule\\
 calculation & Krasnikov et al. \cite{Krasnikov:1982gf} & 
 0.003 &  & Sum rule\\
 & Ioffe and Smilga \cite{Ioffe:1983ju} & 0.006 &  & Sum rule\\
 & Tomozawa \cite{Tomozawa:1980rc} & 0.006 &  & Quark model\\
 & Brodsky et al. \cite{Brodsky:1983st} & 0.03 &  & \\
 \hline
 & Hara et al. \cite{Hara:1986hk} & 0.03 & & WF, $a=0.11$ fm\\
 & Bowler et al. \cite{Bowler:1987us} & 0.013 & 0.010 & WF, $a=0.22$ fm\\
 Lattice QCD & Gavela et al. \cite{Gavela:1988cp} & 0.0056(8) & $\simeq |\alpha|$ &
 WF, $a=0.09$ fm\\
 $N_f=0$ &JLQCD \cite{Aoki:1999tw} & 0.015(1) & 0.014(1) & WF, $a=0.09$ fm\\
 & CP-PACS \& JLQCD \cite{Tsutsui:2004qc} & 0.0090(09)($^{+5}_{-19}$) &
 0.0096(09)($^{+6}_{-20}$) & WF, continuum limit\\
 & Aoki et al. \cite{Aoki:2006ib} & 0.0100(19) & 0.0108(21) & DWF, $a=0.15$ fm\\
 \hline
 \begin{tabular}{c}
  Lattice QCD\\
  $N_f=2$\\
 \end{tabular} & Aoki et al. \cite{Aoki:2006ib} & 0.0118(21) & 0.0118(21) & DWF,
 $a=0.12$ fm\\
 \hline
 \begin{tabular}{c}
  Lattice QCD\\
  $N_f=2+1$\\
 \end{tabular} & This work & 0.0112(25) & 0.0120(26) & DWF, 
 $a=0.12$ fm\\
 \hline
 \hline
\end{tabular}
\caption{Comparison of the low energy parameter of the nucleon decay 
 chiral Lagrangian $\alpha$ and $\beta$ among various QCD model
 calculation, lattice results  in the literatures and the results from
 this work. In lattice QCD calculations, WF and DWF mean Wilson and domain-wall fermions.
 The results for $N_f = 2$, and our results for $N_f = 2+1$ are shown with the total error consisting
 of statistical and systematic errors on the bare matrix element and
 renormalization constant. The errors on the results from $N_f = 0$ are only statistical.
 }
\end{table}

%% file: lifetimes.tex
\begin{table}[htb]
\begin{center}
\begin{tabular}{rl|c|c}
\hline
\multicolumn{2}{c|}{Decay Mode}      & Lifetime bound(yrs)   & $A_{\mathrm{GUT}} $ bound $(M_{\mathrm{GUT}}^{-4})$\\
\hline\hline
$p\rightarrow$ & $e^+\pi^0$     & $> 8.2\times10^{33}$  & $<44$\\
\hline
$p\rightarrow$ & $e^+\pi^0$      & $> 8.2\times10^{33}$  & $<37$\\
$p\rightarrow$ & $K^+\bar{\nu}$      & $> 2.3\times10^{33}$  & $<76$\\
$n\rightarrow$ & $K^0\bar{\nu}$      & $> 1.3\times10^{32}$  & $<733$\\
\hline
\end{tabular}
\end{center}
\caption{The experimental proton partial lifetime bounds at $90\%$ CL from \cite{Nishino:2008, Kobayashi:2005pe} and the bound on $A_{\mathrm{GUT}}$ at a 68\% CL that this lifetime bound implies. This bound is given in units of $M_{\mathrm{GUT}}^{-4}$, the numbers quoted in the table are therefore dimensionless. The first line is for a decay mediated by a heavy gauge boson, the second and subsequent lines are for decays mediated by a colour triplet Higgs.}
\label{tab:lifetimes}
\end{table}